\documentclass[showpacs,pre,amssymb,amsmath,preprint]{revtex4}
\pdfoutput=1
\usepackage{bm}
\usepackage{graphics}
\usepackage{multirow}		
\usepackage{rotating}		
\usepackage{floatflt}
\usepackage{color}
\numberwithin{equation}{section}

\usepackage{textcomp}		

\usepackage{psfrag}
\usepackage{amsmath}
\usepackage{amssymb}
\usepackage{textcomp}		
\usepackage{array}		
\usepackage{multirow}		
\usepackage{rotating}		
\usepackage{nag,hyperref}

\def \dd{\hbox{d}}

\def \dd{\hbox{d}}

\newcommand{\br}{\mathbf{r}}

\newcommand{\bq}{\mathbf{q}}
\newcommand{\bk}{\mathbf{k}}
\newcommand{\bl}{\mathbf{l}}

\newcommand{\be}{\begin{equation}}
\newcommand{\ee}{\end{equation}}
\newcommand{\bea}{\begin{eqnarray}}
\newcommand{\eea}{\end{eqnarray}}
\newcommand{\la}{\label}

\newcommand{\bx}{\mathbf{x}}

\begin{document}

\title{HARTREE-FOCK  ANALYSIS OF THE EFFECTS OF LONG-RANGE INTERACTIONS ON THE BOSE-EINSTEIN CONDENSATION}
\author{A. Alastuey*, J. Piasecki** and P. Szymczak**}
\affiliation{* Univ Lyon, ENS de Lyon, Univ Claude Bernard Lyon 1, CNRS, Laboratoire de Physique, F-69342 Lyon, France \\ 
**Institute of Theoretical Physics, Faculty of Physics, 
University of Warsaw,  Pasteura 5, 02-093 Warsaw, Poland}
\date{\today}
\begin{abstract}

We consider a Bose gas with two-body Kac-like scaled interactions $V_{\gamma}(r)=\gamma^3 v(\gamma r)$ where $v(x)$ is a given repulsive and integrable 
potential, while $\gamma$ is a positive parameter which controls the range of the interactions and their amplitude at a distance r.
Using the Hartree-Fock approximation we find that, at finite non-zero temperatures, the Bose-Einstein condensation is destroyed by 
the repulsive interactions when they are sufficiently long-range. More 
precisely, we show that for $\gamma$ sufficiently small but finite the off-diagonal part 
of the one-body density matrix always vanishes at large distances. Our analysis sheds light 
on the coupling between critical correlations and long-range interactions, 
which might lead to the breakdown of the off-diagonal long-range order even beyond the Hartree-Fock approximation. 
Furthermore, our Hartree-Fock analysis shows the existence of a threshold value $\gamma_0$ above which the 
Bose-Einstein condensation is restored. Since $\gamma_0$ is an unbounded increasing function of the temperature this implies
for a fixed $\gamma$, namely for a fixed scaled potential, that a condensate cannot form above some critical temperature whatever the value of the density.
\end{abstract}


\maketitle

\section{Introduction}

One century ago the Bose-Einstein (BE) condensation was first introduced~\cite{BEoriginal} for an ideal gas of 
identical bosons enclosed in a box in three dimensions. 
In the thermodynamic limit the BE condensation was predicted to occur above 
some critical density depending on the temperature (see e.g. the thorough review~\cite{ZUK1977} or the textbook~\cite{PS2003}).
Paradoxically, this phase transition is a pure effect of Bose statistics as no particle interactions are needed. On the contrary, 
such interactions might even destroy the condensation. The present paper is motivated by the fundamental question of the status of the 
BE condensation in presence of particle interactions for a homogeneous infinite system~\cite{S2010}. No trapping external potential is present, like 
in laboratory realizations with cold atoms (see the review~\cite{ColdAtoms}). Note that recent experiments~\cite{ExperimentFlat} 
where the trapping potential is almost flat, 
should be more relevant for the homogeneous infinite system as considered here.  

In this paper we study a three-dimensional Bose gas with two-body Kac-like scaled interactions $V_{\gamma}(r) = \gamma^3 v(\gamma r)$, where 
$v(x)$ is a given positive repulsive potential such that its integral over the whole space is equal to $a$. The 
positive parameter $\gamma$ controls both the range and the amplitude of $V_{\gamma}(r)$.
In the presence of interactions, the status of the BE condensation must be discussed in relation with 
the existence of Off-Diagonal Long Range Order (ODLRO) for the one-body density matrix,
which in this case is the relevant order parameter~\cite{Penrose1951,Penrose1956,Yang1962}.  
Here, we address this rather challenging question at finite non-zero temperatures within 
the Hartree-Fock (HF) approximation as a first step. Naturally, one needs to remember that such mean-field theory 
does not describe the exact behavior of the system in the critical region. It is well known that the HF approximation 
suffers from various drawbacks, in particular because of its poor description of the condensate~\footnote{The HF description 
of the condensate can be improved within the so-called Hartree-Fock-Bogoliubov (HFB) 
approximation exposed in \textit{e.g.} the textbook~\cite{HFB} or Ref.~\cite{HFBbis}, which
incorporates the Bogoliubov description of the condensate.}. Nevertheless, and similarly to the spirit of other works (see \textsl{e.g.} Ref.~\cite{OS1997}), 
we can reasonably expect that the mean-field HF theory would provide important insights into the main mechanisms at work.

Although the HF approximation was introduced long ago~\cite{HartreeFock} and widely applied to an interacting Bose 
gas (see e.g. the textbooks~\cite{FW1971,PS2003,MR2004}), to our knowledge the 
possible influence of the potential range on critical properties has never been 
studied in detail. It is worth noticing that the HF equations can be solved in two extreme cases 
$\gamma=0$ and $\gamma=\infty$. 
It turns out that the HF prediction for $\gamma=0$ becomes identical to the exact result proved for the so-called 
imperfect Bose gas or mean-field model~\cite{BLS1984,PZ2004}, where the particles interact \textit{via} a constant potential 
$a \rho$ with $a >0$ and $\rho$ the particle density. Then the eigenstates of the corresponding mean-field 
Hamiltonian reduce to symmetric products of one-body states
while the kinetic particle energies are shifted by the constant $a \rho$. Not surprisingly, 
the BE condensation is preserved at the same critical density as in the ideal case~(see the review \cite{ZB2001} and references given therein). 
At $\gamma=\infty$, the HF approximation 
provides a simple shift of the particle kinetic energies by the constant $2 a \rho$, so the BE condensation persists. For a 
sufficiently short-range potential, namely here for $\gamma$ sufficiently large, 
the HF equations can be solved within a perturbative scheme around the solution obtained
for the delta-potential $V_\infty(\br)=a\delta(\br)$~\cite{FW1971}. 
This leads to the prediction that the BE condensation persists for $\gamma$ large but finite, with some shift of the critical temperature at fixed 
density~\cite{L1962,FW1971,PS2003}. Here we argue that such perturbative analysis fails for $\gamma$ sufficiently small. In other words, 
the HF one-body density matrix obtained through a fully self-consistent solution of the HF equations does not exhibit ODLRO 
for $\gamma$ sufficiently small.

Interestingly, our analysis shows that the point $\gamma=0$ is singular, in the sense that the critical point is erased as soon as 
$\gamma$ is small but finite. Indeed, for $\gamma > 0$, the situation 
is quite different from $\gamma=0$, because the two-body interactions $V_{\gamma}(r)$ do not reduce to a constant, hence
the previous simple picture where particles feel a constant potential breaks down. In fact, since
$V_{\gamma}(r)$ varies over large length scales of order $\gamma^{-1}$, 
it interferes with the slowly decaying critical tails which emerge when the BE condensation takes place. For large values of $\gamma$,  
the BE condensation persists because the corresponding short-range particle interactions 
can be safely neglected at large distances where critical correlations take place. 

The paper is organized as follows. In Section~\ref{Model}, we define the model and we 
recall the integral equations which define the HF approximation. We proceed to an asymptotic analysis of the HF equations for 
small values of $\gamma$ in Section~\ref{AbsenceODLRO}. Exploiting the quasi-delta nature in Fourier space of the potential 
$V_{\gamma}(r)$, and assuming \textit{a priori} slower variations of the HF effective potential, we infer 
the asymptotic small-$\gamma$ form of these equations showing that they rule out the possibility of ODLRO 
at sufficiently small but finite $\gamma$. These asymptotic equations play a central role in the derivation of the small-$\gamma$ 
expansions of the effective HF potential and of the density. The corresponding explicit calculations 
whose details are given in Appendix~\ref{ASEXP} (leading terms) and Appendix~\ref{SubleadingCorrections} (subleading corrections) 
are performed in a fully consistent way. 
The resulting effective potential displays indeed slow variations, which justifies \textit{a posteriori} the validity of our derivations. This confirms 
the lack of ODLRO at sufficiently small but finite $\gamma$. 
Taking into account the persistence of the ODLRO for sufficiently large values of 
$\gamma$ we then infer the existence of a threshold value $\gamma_0$ above which the BE condensation is restored.
 
The above predictions are confirmed in Section~\ref{NR} by the numerical solution of the HF equations for a Gaussian potential $v(x)$. 
Moreover, the behavior of $\gamma_0$ with respect to the temperature 
is rather accurately reproduced by a simplified effective-mass version of the HF equations
discussed in Appendix~\ref{EffectiveMass}. Interestingly, such behavior 
immediately implies the existence of a critical temperature above which the BE condensation cannot occur.  
Eventually, concluding comments are given in Section~\ref{Conclusion}.

\section{Model and definitions} 
\la{Model} 

\subsection{Bose gas with Kac-like two-body interactions }
\la{BoseGas}

We consider a model of $N$ non-relativistic spinless bosons with mass $m$,
described by the Hamiltonian
\be
\la{hamscaled}
{\cal H}_{\gamma}=  -\sum_{i=1}^N \frac{\hbar^2}{2m}\Delta_{\br_i} + \sum_{i<j=1}^{N}V_{\gamma}(|\br_{i}-\br_{j}|) \; ,
\ee
where the two-body interactions $V_{\gamma}(r)$ depend on a scaling parameter 
$\gamma >0$  
\be
\la{Vscaled}
V_{\gamma}(r) = \gamma^3 v(\gamma r) \; .
\ee
Here, the spherically symmetric pair potential $v(r)$ is assumed to be positive and integrable.  
The dimensionless parameter $\gamma$ controls the range of the potential $V_{\gamma}(r)$ and its amplitude. When 
$\gamma$ is varied while $v(r)$ is kept fixed, the mean-field energy defined by
the integral
\be
\la{aconstant}
\int \, \dd \br \, V_{\gamma}(r) = 4\pi \int_{0}^{\infty}\dd r \, r^2 v(r) = a > 0
\ee
remains constant. 

We start by enclosing the particles in a finite box with volume $\Lambda$.  
We describe an equilibrium state of that system in the grand canonical 
ensemble characterized by temperature $T$ and chemical potential $\mu$. Then, the 
thermodynamic limit (TL), $\Lambda \to \infty$ at fixed $T$ and fixed $\mu$, 
is taken, and it is assumed to exist. We consider values of $(T,\mu)$ such that
the resulting infinite system is in a fluid state, 
invariant under translations and rotations. In the following, all local quantities specific to that infinite system 
are implicitly defined through the usual TL procedure. For the sake of simplicity in the notations, we 
assume that the TL has been taken once for all. In particular, all involved spatial integrals run over the whole space.

\subsection{Hartree-Fock approximation} 
\la{HFapproximation}

The well-known Hartree-Fock approximation is based on two non-linear integral equations which couple the  
particle density $\rho$ with an effective potential $\hat{\phi}_\gamma$ in Fourier space.  
If the effective energy 
\begin{equation}
\la{effectiveenergy}
U_{\gamma}(\bk)=\epsilon(\mathbf{k}) -\mu + a\rho + \hat{\phi}_\gamma(\mathbf{k}) \;  
\end{equation}
with $\epsilon(\mathbf{k})=\hbar^2 k^2/(2m)$,
is strictly positive for any $\mathbf{k}$, then 
the HF equations read
\begin{equation}
\la{HFdensity}
\rho=\frac{1}{(2\pi)^3} \int \dd \mathbf{k} 
\frac{1}{\exp[\beta (\epsilon(\mathbf{k})-\mu + a\rho + \hat{\phi}_\gamma(\mathbf{k}))]-1 } \; 
\end{equation} 
and
\begin{equation}
\la{HFpotential}
\hat{\phi}_\gamma(\mathbf{q})=\frac{1}{(2\pi)^3} \int \dd \mathbf{k} 
\frac{\hat{V}_\gamma(\mathbf{q} - \mathbf{k})}
{\exp[\beta (\epsilon(\mathbf{k})-\mu + a\rho + \hat{\phi}_\gamma(\mathbf{k}))]-1 } \; .
\end{equation} 
The system is then predicted to be in a normal phase, with a fast decay of the off-diagonal 
matrix element $\langle \mathbf{r}_2|{\cal D}^{(1)}|\mathbf{r}_1 \rangle$ of the one-body 
density matrix when $|\br_2-\br_1| \to \infty$.

At finite non-zero temperatures ($\beta = 1/k_{\rm B}T < \infty$) the emergence of the Bose-Einstein condensation 
is signaled by vanishing of the effective 
energy~(\ref{effectiveenergy}) at $\bk=\mathbf{0}$, namely 
\begin{equation}
\la{ODLROcondition}
\hat{\phi}_\gamma(\mathbf{0}) + a\rho -\mu=0 \; . 
\end{equation}
At the corresponding critical point, the critical density $\rho_{\rm cri}$ 
and the critical effective potential $\hat{\phi}_{\rm cri}$ are still given 
by Eqs.~(\ref{HFdensity}) and~(\ref{HFpotential}). However, the off-diagonal 
matrix element $\langle \mathbf{r}_2|{\cal D}^{(1)}|\mathbf{r}_1 \rangle$ decays now slowly $\sim |\br_2-\br_1|^{-1}$
at large distances which marks the onset of ODLRO. Above the critical density, for  $\rho >\rho_{\rm cri}$, a 
condensate with density $\rho_{\rm cond} > 0$ emerges, while the ODLRO condition~(\ref{ODLROcondition}) is always satisfied. Then 
$\langle \mathbf{r}_2|{\cal D}^{(1)}|\mathbf{r}_1 \rangle$ tends to $\rho_{\rm cond}$ when $|\br_2-\br_1| \to \infty$, while 
the difference $\langle \mathbf{r}_2|{\cal D}^{(1)}|\mathbf{r}_1 \rangle - \rho_{\rm cond}$ still decays slowly $\sim |\br_2-\br_1|^{-1}$. 
The corresponding equations relating the total density to the effective potential then read 
\be
\la{DensCond}
\rho = \rho_{\rm cond} + \frac{1}{(2\pi)^3} \int \dd \mathbf{k} 
\frac{1}{\exp[\beta (\epsilon(\mathbf{k}) +\hat{\phi}_\gamma(\mathbf{k})-\hat{\phi}_\gamma(\mathbf{0}) )]-1 }  
\ee
and
\begin{equation}
\la{PhiCond}
\hat{\phi}_\gamma(\mathbf{q})= \rho_{\rm cond} \hat{V}_\gamma(\mathbf{q}) + \frac{1}{(2\pi)^3} \int \dd \mathbf{k} 
\frac{\hat{V}_\gamma(\mathbf{q} - \mathbf{k})}
{\exp[\beta (\epsilon(\mathbf{k})  + \hat{\phi}_\gamma(\mathbf{k})-\hat{\phi}_\gamma(\mathbf{0}))]-1 } \; .
\end{equation}

Although the onset of ODLRO is characterized by the simple condition~(\ref{ODLROcondition}), the search for the 
corresponding critical point remains quite a hard task because both the density and effective potential are 
related by non-linear integral equations which are too difficult to be solved analytically in general. However, 
these equations become significantly simpler in two limiting cases $\gamma=0$ and $\gamma=\infty$. For $\gamma=0$, 
$\phi_0$ identically vanishes and $\rho$ reduces to its ideal expression with the shifted chemical 
potential $(\mu-a \rho)$. We then recover the exact result for the imperfect Bose gas with 
constant interactions~\cite{BLS1984,PZ2004}: the system undergoes the Bose-Einstein condensation at a critical density 
which coincides with its ideal counterpart, $\rho_{0,{\rm cri}}=\rho_{\rm cri}^{(\rm id)}$, while the corresponding critical potential 
is $\mu_{0,{\rm cri}}=a\rho_{\rm cri}^{(\rm id)}$. The case $\gamma=\infty$ is associated with the delta-potential 
$V_\infty(\br)=a\delta(\br)$, for which $\hat{\phi}_\infty(\mathbf{q})$ reduces to the constant $a \rho$. The situation then becomes 
analogous to that of $\gamma=0$, with a shifted chemical potential which is now $(\mu-2 a \rho)$, 
so the Bose-Einstein condensation persists 
at the same critical density $\rho_{\infty,{\rm cri}}=\rho_{\rm cri}^{(\rm id)}$. For $0  < \gamma < \infty$, which corresponds to
potentials $V_\gamma(r)$ with finite range, similar simple results are not available since $\phi_\gamma(\br)$ 
no longer reduces to a constant.  

Finally, some simple analytical properties of the HF quantities at $\gamma$ finite can be derived as follows. 
According to the Kac scaling~(\ref{Vscaled}), the Fourier transform $\hat{V}_\gamma(\mathbf{k})$ reduces to $\hat{v}(\mathbf{k}/\gamma)$
where $\hat{v}(\mathbf{l})$ is the Fourier transform with respect to $\mathbf{x}$ of $v(\mathbf{x})$ 
which does not depend on $\gamma$. Since $|\hat{v}(l)| \leq \hat{v}(0)=a$, the expressions~(\ref{HFpotential}) and~(\ref{PhiCond}) for the effective 
potential are readily bounded from above. We find the simple inequality
\be
\la{BoundPhi}
|\hat{\phi}_\gamma(\mathbf{q})| \leq a \rho \;  
\ee
which holds in any phase. Furthermore, for potentials such that $0 \leq \hat{v}(l) \leq a$ for any $l$, 
$\hat{\phi}_\gamma(\mathbf{q})$ is positive for any $q$, so the inequality~(\ref{BoundPhi}) becomes
\be
\la{BoundPhibis}
0 \leq \hat{\phi}_\gamma(\mathbf{q}) \leq a \rho \; .
\ee
Now, if a condensate emerges at some critical chemical potential 
$\mu_{\gamma,{\rm cri}}$ and critical density $\rho_{\gamma,{\rm cri}}$,
the ODLRO condition~(\ref{ODLROcondition}) is satisfied for any $\mu \geq \mu_{\gamma,{\rm cri}}$ with the corresponding density  
$\rho_{\gamma}(\mu,\beta)$. Combining this condition with the inequality~(\ref{BoundPhibis}) for $q=0$,
we then obtain
\be
\la{BoundsEOS}
a \rho_{\gamma}(\mu,\beta) \leq \mu \leq 2 a \rho_{\gamma}(\mu,\beta) \; ,
\ee
Hence, the curve describing the equation of state $\rho=\rho_{\gamma}(\mu,\beta)$ in the $(\mu,\rho)$-plane 
for $\mu \geq \mu_{\gamma,{\rm cri}}$, lies in the wedge defined by the straight lines 
$\rho=\mu/a$ and $\rho=\mu/(2a)$. Inequalities~(\ref{BoundPhibis}) and~(\ref{BoundsEOS}) can serve as checks 
for numerical calculations, as illustrated in Section~\ref{NR} for a Gaussian potential
$v(\mathbf{x})$.

\section{ABSENCE OF  ODLRO FOR SUFFICIENTLY LARGE BUT FINITE RANGE OF THE POTENTIAL }
\la{AbsenceODLRO}

In this Section, we consider the case of small values of $\gamma$, namely long-range interactions. We fix the temperature at 
a non-zero value ($\beta < \infty$), as well as the chemical potential $\mu$, and we seek for 
the behavior of various quantities of interest 
when $\gamma \to 0$. For the sake of notational convenience, we do not write explicitly the dependence on $(\mu,\beta)$ 
of the effective potential and of the effective energy which are denoted by 
$\hat{\phi}_\gamma(\mathbf{k})$ and $U_\gamma(\bk)$, respectively, while we keep all  arguments specified in the density $\rho=\rho_{\gamma}(\mu,\beta)$. We 
assume that both $\hat{\phi}_\gamma(\mathbf{k})$ and  
$\rho_{\gamma}(\mu,\beta)$ are continuous functions of $\gamma$ at $\gamma=0$ for fixed values of their
respective arguments $\mathbf{k}$ and $(\mu,\beta)$. This means that $\hat{\phi}_\gamma(\mathbf{k})$ vanishes when 
$\gamma \to 0$, while $\rho_{\gamma}(\mu,\beta)$ approaches $\rho_{0}(\mu,\beta)$ which is nothing else but the density of
the mean-field model. 
 
\subsection{Slow variations of the effective potential and the asymptotic HF equations}
\la{SVAS}

For our purpose, it is convenient to rewrite
the integral equation~(\ref{HFpotential}) for $\hat{\phi}_\gamma$ as 
\begin{equation}
\la{HFpotentialBis}
\hat{\phi}_\gamma(\mathbf{q})=\frac{1}{(2\pi)^3} \int \dd \mathbf{k} 
\frac{\hat{v}((\mathbf{q} - \mathbf{k})/\gamma)}
{\exp[\beta (\epsilon(\mathbf{k})-\mu + a \rho_{\gamma}(\mu,\beta) + \hat{\phi}_\gamma(\mathbf{k}))]-1 } \; .
\end{equation} 
The integral over $\bk$ in the r.h.s. of this equation has to be performed on the product of  
function  $\hat{v}((\mathbf{q} - \mathbf{k})/\gamma)$ varying fast 
around $\bk=\bq$ over a scale $\gamma$, times the effective Bose distribution
\be
\la{BEdistribution}
n_\gamma^{\rm B}(\bk)= \frac{1}
{\exp[\beta U_\gamma(\bk) ]-1 } \; .
\ee 
Accordingly, the leading contributions  
are expected to arise in a relatively small region where $\bk$ is close to $\bq$, namely $\bk=\bq + \gamma \bl$ with $\bl$ finite. It is the useful to recast 
the integral equation~(\ref{HFpotentialBis}) as 
\begin{equation}
\la{HFpotentialTer}
\hat{\phi}_\gamma(\mathbf{q})=\frac{\gamma^3}{(2\pi)^3} \int \dd \mathbf{l} \; \hat{v}((\mathbf{l} )
n_\gamma^{\rm B}(\bq + \gamma \bl) \; .
\end{equation} 
Let us assume \textsl{a priori} that $n_\gamma^{\rm B}(\bq + \gamma \bl)$ varies around $n_\gamma^{\rm B}(\bq)$ on a slower scale than $\gamma$. 
Since $|\mathbf{l}|$  cannot exceed the range of the potential $\hat{v}$, we can 
replace $n_\gamma^{\rm B}(\bq + \gamma \bl)$ by its Taylor expansion in powers of $\gamma \bl$ around $n_\gamma^{\rm B}(\bq)$. Keeping only the
leading order term, reduces the integral equation~(\ref{HFpotentialTer}) to the local equation
\begin{equation}
\la{HFas}
\hat{\phi}_\gamma(\bq)= \gamma^3 v(0) n_\gamma^{\rm B}(\bq)  = \frac{\gamma^3 v(0) }
{{\exp[\beta (\epsilon(\mathbf{q})+ \eta_{\gamma}(\mu,\beta) + \hat{\phi}_\gamma(\mathbf{q}))]-1 } } \; ,
\end{equation}
where we set
\be
\la{eta}
\eta_{\gamma}(\mu,\beta)= a\rho_{\gamma}(\mu,\beta) -\mu \; .
\ee

Equations~(\ref{HFas}) and (\ref{eta}) have to be solved self-consistently together with equation~(\ref{HFdensity}) which provides
the density $\rho_{\gamma}(\mu,\beta)$ in terms of $\hat{\phi}_\gamma(\bq)$. They play a central role 
in the derivation of the small-$\gamma$ expansions of the HF quantities of interest, because the resulting 
effective potential $\hat{\phi}_\gamma(q)$ does vary on a scale slower than $\gamma$ for any $q$, so the \textit{a priori} assumption leading to the 
local equation~(\ref{HFas}) is justified\textit{ a posteriori}. Hence, as described in Section~(\ref{SmallExpansion}), beyond the leading terms, 
subleading corrections can be consistently computed, in particular those 
arising from the Taylor expansion of $n_\gamma^{\rm B}(\bq + \gamma \bl)$ around $n_\gamma^{\rm B}(\bq)$ in powers of 
$\gamma \bl$ in the integral HF equation~(\ref{HFpotentialTer}).

Before turning to the explicit calculations of the small-$\gamma$ expansions, a 
simple and fundamental consequence of the HF asymptotic equation~(\ref{HFas}) must be emphasized. 
Let us rewrite equation~(\ref{HFas}) as
\be
\label{HFasBis}
\hat{\phi}_\gamma(\bq)\left\{ \exp[\beta (\epsilon(\mathbf{q}) + \eta_{\gamma}(\mu,\beta) + \hat{\phi}_\gamma(\mathbf{\bq}))]-1 \right\}
= \gamma^3 v(0) \; .
\ee
The right hand side does not depend on $q$. We thus also have
\be
\label{HFasTer}
\hat{\phi}_\gamma(\mathbf{0}) \left\{ \exp[\beta (\hat{\phi}_\gamma(\mathbf{0}) + \eta_{\gamma}(\mu,\beta) )]-1 \right\}
= \gamma^3 v(0) \; .
\ee
Hence, any $\gamma >0$ excludes the possibility of satisfying the ODLRO condition~(\ref{ODLROcondition}). 
This strongly suggests that within the HF approximation, the BE condensation is 
indeed removed for $\gamma$ sufficiently small. In Section~\ref{AbsenceBE} we comment and interpret this important result.

Finally, let us mention some other simple properties of the asymptotic HF equation~(\ref{HFas}). First, the identity,   
\begin{equation}
\la{HFpotentialIdentity}
\int \dd \mathbf{q} \hat{\phi}_\gamma(\mathbf{q})= (2\pi)^3 \gamma^3 v(0) \rho  \; ,
\end{equation}
which follows by integrating over $\bq$ both sides of the HF equation~(\ref{HFpotential}), 
is still fulfilled by the solution of the asymptotic HF equation~(\ref{HFas}). Moreover
equation~(\ref{HFasBis}) implies that  $\hat{\phi}_\gamma(\bq)=\hat{\phi}_\gamma(q)$ is a decreasing function
of $q$, whereas the effective energy~(\ref{effectiveenergy}) 
is monotonically increasing.

\subsection{Small-$\gamma$ expansions of the HF effective potential and density}
\label{SmallExpansion}

The small-$\gamma$ expansions are calculated in Appendix~\ref{ASEXP} (leading terms) and Appendix~\ref{SubleadingCorrections} 
(subleading corrections) for the three different regions which naturally emerge, namely $\mu$ below the critical mean-field 
chemical potential $\mu_{0,{\rm cri}}$, $\mu$ close to $\mu_{0,{\rm cri}}$ and $\mu$ above $\mu_{0,{\rm cri}}$.
Here we sketch the main steps of the derivations, and we give the small-$\gamma$ expansions of 
$\rho_{\gamma}(\mu,\beta)-\rho_{0}(\mu,\beta)$ and of the ODLRO parameter 
\begin{equation}
\la{ODLROparameter}
U_\gamma(0)=\hat{\phi}_\gamma(\mathbf{0}) + \eta_{\gamma}(\mu,\beta) \; .
\end{equation}

For $\mu < \mu_{0,{\rm cri}}$, a straightforward perturbative solution of the local equation~(\ref{HFas}) shows that 
$\hat{\phi}_\gamma(\mathbf{k})$ is of order $\gamma^3$ at leading order (see formula~(\ref{PotBC})), while it varies over a finite scale $O(1)$ much 
larger than $\gamma$. The resulting difference $\rho_{\gamma}(\mu,\beta)-\rho_{0}(\mu,\beta)$ then is also of order 
$\gamma^3$ at leading order (see formula~(\ref{DenBC})). The ODLRO parameter~(\ref{ODLROparameter}) then goes to $\eta_{0}(\mu,\beta)$ 
which is strictly positive. This analysis is no longer valid near the critical mean-field chemical potential 
because the coefficients of the previous $\gamma^3$-terms diverge when $\mu \to \mu_{0,{\rm cri}}$.

For studying the vicinity of the critical point of the mean-field model, we set $\mu= \mu_{0,{\rm cri}} + \nu$ and 
$\rho_{\gamma}(\mu,\beta)=\rho_{\rm cri}^{(\rm id)} + n_{\gamma}(\nu,\beta)$. Moreover we scale $\nu$ as  
$\nu = \nu^\ast \gamma^{3/4}$, and we compute the resulting small-$\gamma$ expansions at fixed $\nu^\ast$. The leading 
contributions to $n_{\gamma}(\nu,\beta)$ arise from small values of $q$, for which we use the effective potential~(\ref{PotNCsmall}) 
inferred from the purely local equation~(\ref{HFas}). 
The corresponding consistency equation for the density is then expressed in terms of the 
function $K$ defined by its integral representation~(\ref{Kfunction}). This provides the small-$\gamma$ 
expansion of $ n_{\gamma}(\nu,\beta)$,
\be
\la{DensDevNCexpansion}
n_{\gamma}(\nu,\beta) = a^{-1} \left[ \gamma^{3/4} \nu^\ast + 
2\gamma^{3/2} \sqrt{\beta^{-1} v(0)} \sinh K^{-1}\left( \pi^2\lambda_{\rm dB}^3 [\beta v(0)]^{-1/4} \nu^\ast /a  \right) + o(\gamma^{3/2})  \right]  \; , 
\ee 
where $K^{-1}$ is the inverse of $K$. We stress that the effective potential $\hat{\phi}_\gamma(\mathbf{k})$  
does vary on a larger scale than $\gamma$ for any $q$, which guarantees the validity of expansion~(\ref{DensDevNCexpansion}) 
up to order $\gamma^{3/2}$ included. In fact, 
the corrections to the local equation~(\ref{HFas}) derived from
the integral equation~(\ref{HFpotentialBis}) provide subleading terms of order $\gamma^2$ in this expansion. For $\mu= \mu_{0,{\rm cri}}$, \textit{i.e.} 
$\nu^\ast=0$, the small-$\gamma$ expansion of the density then reads
\be
\la{DensNCpoint}
\rho_{\gamma}(\mu_{0,{\rm cri}},\beta) = \rho_{\rm cri}^{(\rm id)} + 
2\gamma^{3/2} a^{-1}\sqrt{\beta^{-1} v(0)} \sinh K^{-1}(0) + o(\gamma^{3/2})    \; , 
\ee
with $ K^{-1}(0) < 0$. Note that the emergence of the $\gamma^{3/2}$-correction is consistent with the divergence of the coefficient 
of the $\gamma^3$-correction for $\mu < \mu_{0,{\rm cri}}$ when $\mu \to \mu_{0,{\rm cri}}$. 
As expected from the simple argument presented in Section~\ref{SVAS},
the ODLRO condition~(\ref{ODLROcondition}) is no longer reached in the vicinity of the critical point of the 
mean-field model. The order parameter~(\ref{ODLROparameter}) indeed behaves at leading order as 
\begin{equation}
\la{ODLROparameterNC}
U_{\gamma}(0) \sim \gamma^{3/2} \sqrt{\beta^{-1} v(0)} \exp \left( K^{-1}\left( \pi^2\lambda_{\rm dB}^3 [\beta v(0)]^{-1/4} \nu^\ast /a  \right) \right) \; .
\end{equation}
which remains strictly positive for any $\nu^\ast$. 

Now we fix  $\mu > \mu_{0,{\rm cri}} $, and as above 
we still define $\nu= \mu - \mu_{0,{\rm cri}} $ and $n_{\gamma}(\nu,\beta)= \rho_{\gamma}(\nu,\beta) - \rho_{\rm cri}^{(0)} $. 
For the mean-field model $\eta_{0}(\mu,\beta)=0$ since the corresponding equation of state is $n_{0}(\nu,\beta)=a \nu$ for 
$\nu >0$. Therefore, by continuity $\eta_{\gamma}(\mu,\beta)$ vanishes in the limit $\gamma \to 0$. Like in the vicinity 
of the critical point of the mean-field model, both $\eta_{\gamma}(\nu,\beta)$ and $\hat{\phi}_{\gamma}(q)$ become small 
for $\gamma$ small, so the analysis of the corresponding leading terms can be carried out along similar lines. 
The corresponding consistency equation for $n_{\gamma}(\nu,\beta)$ again involves the function $K$, from which we infer 
\be
\la{DensDevACexpansion}
n_{\gamma}(\nu,\beta) = a^{-1} \left[ \nu - \left( \frac{15 \pi^2 \lambda_{\rm dB}^3  \nu}{2 \sqrt{2} \beta^{3/2} v(0)^{3/2} a} \right)^{2/5} v(0) \gamma^{6/5} 
 + o(\gamma^{6/5})  \right]  \; . 
\ee
The effective potential $\hat{\phi}_\gamma(\mathbf{k})$  
still varies on a larger scale than $\gamma$ for any $q$, so the corrections to the local equation~(\ref{HFas}) ultimately give raise 
to subleading terms in the small-$\gamma$ expansion~(\ref{DensDevACexpansion}), which turn to be of order $o(\gamma^{6/5})$.  
The HF correction to the 
equation of state (EOS) of the mean-field model, namely $n=\nu/a$ for $\nu \geq 0$,
is here of order $\gamma^{6/5} $ for $\nu > 0$, in agreement with the divergence of the $\gamma^{3/2}$-correction 
in the vicinity of the critical point of this model when $\nu^\ast \to \infty$. 
Contrarily to the mean-field model there is no condensate present for $\nu > 0$, since  
the ODLRO parameter~(\ref{ODLROparameter}) behaves as
\be
\la{PotUoriginAC}
U_{\gamma}(0) \sim \left( \frac{2 \sqrt{2} \; a}{15 \pi^2 \lambda_{\rm dB}^3 \beta v(0) \nu} \right)^{2/5} v(0) \; \gamma^{9/5}  \; .
\ee
Nevertheless, the HF density can take arbitrarily high values.

\subsection{Breakdown of the BE condensation}
\label{AbsenceBE}

The main consequence of the previous asymptotic analysis  is the breakdown of the BE condensation for small but finite values of $\gamma$. 
Indeed, in this regime, the order parameter~(\ref{ODLROparameter}) $U_{\gamma}(0)$ never vanishes and remains strictly positive for any 
set $(\mu,\beta)$. In other words the critical point observed for the mean-field model is erased. In fact, 
according to the asymptotic formula~(\ref{PotUoriginAC}), $U_{\gamma}(0)$ vanishes only when $\mu \to \infty$, so the 
critical point can be seen as rejected to $\infty$. This mechanism can be roughly interpreted as follows. 
The slow $1/r$ critical decay of the off-diagonal one-body density matrix at $\gamma=0$ 
which emerges for $r \gg \lambda_{\rm dB}$, becomes drastically altered by contributions of the slow-varying interactions $V_{\gamma}(r)$ 
when their range $\sigma_\gamma \propto 1/\gamma$  becomes much larger than $\lambda_{\rm dB}$.  
In other words, the effects of long-range interactions on critical properties cannot be treated perturbatively, 
and they ultimately eliminate the critical point
for $\gamma$ sufficiently small. At a technical level, this means that such effects must be dealt with through 
a complete self-consistent description of the HF equations. The resulting fractional powers of 
$\gamma$ which control the 
corrections to the EOS of the mean-field model, namely $\gamma^{3/2}$ and $\gamma^{6/5}$ near and above the mean-field critical point 
respectively, are a signature of the singular character of the point $\gamma=0$.

In the opposite limit $\gamma \to \infty$, namely for $\gamma$ sufficiently large, the ODLRO is expected to persist as 
shown by a standard perturbative calculation (see \textit{e.g.} Ref.~\cite{FW1971}). In 
the integral equation~(\ref{HFpotentialBis}) for $\hat{\phi}_{\gamma}(q)$, the almost flat Fourier transform 
$\hat{v}((\mathbf{q} - \mathbf{k})/\gamma)$ of the potential is replaced by its Taylor series expansion  
around $\hat{v}(0)=a$ in powers of $(\mathbf{q} - \mathbf{k})/\gamma$. This generates an expansion of all  
the quantities of interest in integer powers of $1/\gamma^2$ around their respective values for 
the $\delta$-potential $V_\infty(\br)=a\delta(\br)$ where all terms can be straightforwardly computed through a simple recursive scheme. 
The ODLRO condition~(\ref{ODLROcondition}) can then be satisfied order by order, and the $1/\gamma^2$-expansion of the critical density
reads 
\begin{equation}
\label{LGDenCri}
\rho_{\gamma,{\rm cri}}= \rho_{\rm cri}^{(\rm id)} \left[ 1 + \frac{3 \beta a \rho_{\rm cri}^{(\rm id)} \sigma^2}{2 \gamma^2 \lambda_{\rm dB}^2} + O(1/\gamma^4) \right]
\; 
\end{equation}
with the range $\sigma$ of $v(x)$ defined by 
\be
\la{range}
\sigma^2 = \frac{\int \dd \bx \; x^2 \; v(x)}{3 \int \dd \bx \; v(x)}= (3a)^{-1} \int \dd \bx \; x^2 \; v(x) \; ,
\ee
while the ideal critical density given by expression~(\ref{DensIdC}) reduces to 
the well-known formula
\begin{equation}
\label{CriDenIdeal}
\rho_{\rm cri}^{(\rm id)} = \frac{\zeta(3/2)}{(2 \pi \lambda_{\rm dB}^2)^{3/2}}  \; .
\end{equation} 
Assuming the convergence of the $1/\gamma^2$-expansion~(\ref{LGDenCri}) it can be concluded that the ODLRO persists for $\gamma$ 
sufficiently large, namely for a sufficiently short range $\sigma_\gamma = \sigma/\gamma$ 
of the potential $V_{\gamma}(r)$~(see \textit{e.g.} Ref.~\cite{FW1971}). Similarly to the previous interpretation of the 
breakdown of the BE condensation, the potential range $\sigma_\gamma$ becomes much smaller 
than the de Broglie wavelength $\lambda_{\rm dB}$ for $\gamma$ sufficiently large, so 
the $1/r$ critical tails which occur for 
$r \gg \lambda_{\rm dB}$ are only weakly affected by the vanishing interactions $V_{\gamma}(r)$  
at large distances $r \gg \sigma_\gamma$.

The asymptotic results obtained near $\gamma=0$ on the one hand, and near $\gamma=\infty$ on the other hand, imply the existence of  
a threshold value $\gamma_0$ which separates the region where the condensation cannot occur 
($\gamma < \gamma_0$) from the region where it can possibly take place ($\gamma > \gamma_0$). The effects of interactions on critical tails 
depend on both their magnitude and their range. Accordingly, the threshold 
parameter $\gamma_0$ depends on two independent dimensionless parameters, the coupling constant $g=\beta a/\lambda_{\rm dB}^3$ 
which measures the strength of $v(x)$ at the ideal critical density, and the ratio $\lambda=\lambda_{\rm dB}/\sigma$ of 
the de Broglie wavelength to the range of the potential. Although the function $\gamma_0(g,\lambda)$ cannot be explicitly determined 
we can nevertheless reasonably expect that increasing  
$g$ or decreasing $\lambda$, which means increasing the magnitude or the range of the potential, 
makes $\gamma_0$ increase. Hence, for a given potential $v(x)$ with $a$ and $\sigma$ 
fixed once for all, $\gamma_0(T)=\gamma_0(g,\lambda)$ should be an increasing function of the temperature. 
Interestingly, it can be noticed that the typical values $\gamma_{\rm S} $ and 
$\gamma_{\rm L} $  which control 
the convergence of the asymptotic expansions~(\ref{DensNCpoint}) and~(\ref{LGDenCri}) respectively, roughly defined by
equating the first corrections to the leading terms, read
\be
\la{gammaSL} 
\gamma_{\rm S} \propto g^{1/3} \lambda^{-1} \quad \text{and} \quad 
\gamma_{\rm L} \propto  g^{1/2} \lambda^{-1} \; . 
\ee
where we have used $v(0) \propto a/\sigma^3$. 
As functions of $T$, both $\gamma_{\rm S} $ and 
$\gamma_{\rm L} $ reduce to power laws, \textit{i.e. } ${\rm cst} \; T^{2/3}$ and ${\rm cst} \; T^{3/4}$ respectively, 
so they vary from $0$ to $\infty$ when $T$ varies from $0$ to $\infty$. If a similar variation of $\gamma_0(T)$ 
is plausible, its low- and high-temperature behaviors may involve other powers of $T$ than those displayed in $\gamma_{\rm S} $ and 
$\gamma_{\rm L} $ .

The existence of $\gamma_0$ will be confirmed 
in the next Section through the numerical solution of the Hartree-Fock equations in the specific case of a Gaussian potential. 
In this case, we also provide an analytical estimation of $\gamma_0$   
within a simplified version of the Hartree-Fock equations relying on the introduction of an effective mass (see Appendix~\ref{EffectiveMass}). 
Both numerical and analytical results show that the variations of $\gamma_0(T)$ with $T$ are indeed the ones predicted above.

\section{Numerical calculations for a Gaussian potential}
\label{NR}

Now we turn to a numerical solution of the Hartree-Fock equations. 
The potential $v(r)$ involved in Kac scaling is choosen as a Gaussian
$v(r)=(2 \pi \sigma^2)^{-3/2} a \exp(-r^2/(2 \sigma^2))$. Since the Fourier transform of $v(r)$ is still a Gaussian, namely 
$\hat{v}(\bk)=a \exp(-\sigma^2 k^2/2)$, the integrations over the angles $\Omega_{\bk}$ of $\bk$ in the 
integral equation~(\ref{HFpotentialBis}) can be readily performed yielding 
\be
\la{AngularIntegral}
\int \dd \Omega_{\bk}\hat{v}((\bk-\bq)/\gamma) = \frac{2 \pi a \gamma^2}{\sigma^2 kq} 
\left[ \exp(-\sigma^2(k-q)^2/(2\gamma^2))- \exp(-\sigma^2(k+q)^2/(2\gamma^2)) \right] \; .
\ee
The corresponding three-dimensional integral 
is then reduced to a one-dimensional integral over $k$, a particularly useful simplification for numerical 
purposes. After rewriting the full set of Hartree-Fock equations in terms of suitably defined dimensionless quantities 
we proceed to a numerical solution of the coupled equations providing both the effective potential and the density 
as functions of the chemical potential along a given isotherm. We first consider a given fixed temperature, as well as a fixed 
Gaussian potential $v(r)$, while we vary the range $\gamma$. As expected from Section~\ref{AbsenceODLRO} a threshold 
value $\gamma_0$ emerges such that for $\gamma < \gamma_0$ the BE condensate is destroyed. We also analyze the function 
$\gamma_0(T)$ introduced in Section~\ref{AbsenceODLRO}. From such analysis we 
infer that for a given potential 
with fixed range and amplitude there is no BE condensation above some critical temperature.   

\subsection{The coupled equations for dimensionless quantities}
\la{CEDQ}

Now we set $A=\beta (2 \pi \sigma^2)^{-3/2} a $ for the amplitude of the potential in 
$k_{\rm B}T$ units, namely $A=g \lambda^3$ in terms of the dimensionless parameters $g$ and $\lambda$
introduced in Section~\ref{AbsenceODLRO}. Moreover, we can rewrite all the quantities of interest 
in terms of their dimensionless counterparts: 
the dimensionless density $\rho$ now stands for $\rho/\rho_0$ with $\rho_0=(2 \pi \sigma^2)^{-3/2}$, 
the dimensionless chemical potential $\mu$ is in units of $k_{\rm B}T$, while all wavenumbers become dimensionless 
and are measured in units of $\sigma^{-1}$. Eventually, the effective potential $\hat{\phi}_\gamma$ in Fourier space 
is rescaled, which leads to the dimensionless potential $\hat{\varphi}_{\gamma}$ 
linked with the original one by the relation 
$\beta \hat{\phi}_\gamma= A \hat{\varphi}_\gamma $.

The coupled Hartree-Fock equations can be straightforwardly recast in terms of the dimensionless quantities 
and dimensionless wavenumbers. Using the angular integration~(\ref{AngularIntegral}) we find that
the integral equation~(\ref{HFpotentialBis}) becomes
\begin{equation}
\label{Pot}
\hat{\varphi}_\gamma (q)= \frac{\gamma^2}{\sqrt{2\pi}} \int_0^\infty \dd k \; \frac{k}{q} 
\left[ \exp(-(q-k)^2/(2\gamma^2)) - \exp(-(q+k)^2/(2\gamma^2))   \right] n_{\rm B}(k) \; ,
\end{equation}
while the density is given by
\be
\la{Dens}
\rho= \sqrt{ \frac{2}{\pi}} \int_0^\infty \dd k \;  k^2 n_\gamma^{\rm B}(k)
\ee
with the BE distribution
\be
\la{BE}
n_\gamma^{\rm B}(k)=\frac{1}{\exp(\lambda^2 k^2/2 +A \hat{\varphi}_\gamma (k) +A \rho -\mu) -1} \; .
\ee
The above equations are valid in the normal phase.

Along a given isotherm, when $\mu$ increases, a critical point may emerge for some critical 
chemical potential $\mu_{\gamma,{\rm cri}}$, such that the ODLRO condition~(\ref{ODLROcondition}) is satisfied, namely 
\be
\la{ODLROcri}
\mu_{\gamma,{\rm cri}}= A \rho_{\gamma,{\rm cri}} + A \hat{\varphi}_{\gamma,{\rm cri}} (0) \; 
\ee
in terms of the dimensionless quantities. The corresponding critical 
potential is the solution of the integral equation
\begin{multline}
\label{PotCri}
\hat{\varphi}_{\gamma,{\rm cri}} (q)= \frac{\gamma^2}{\sqrt{2\pi}} \int_0^\infty \dd k \; \frac{k}{q} 
\left[ \exp(-(q-k)^2/(2\gamma^2)) - \exp(-(q+k)^2/(2\gamma^2))   \right] \\
\times \frac{1}{\exp(\lambda^2 k^2/2 +A (\hat{\varphi}_{\gamma,{\rm cri}} (k) -\hat{\varphi}_{\gamma,{\rm cri}}  (0) )) -1} \; .
\end{multline}
Interestingly, this equation does not involve neither $\rho_{\gamma,{\rm cri}}$ 
nor $\mu_{\gamma,{\rm cri}}$, so
$\hat{\varphi}_{\gamma,{\rm cri}}(q)$ can first be computed. The critical density then follows as 
\be
\la{DensCri}
\rho_{\gamma,{\rm cri}} = \sqrt{ \frac{2}{\pi}}  \int_0^\infty \dd k \;  k^2 
\frac{1}{\exp(\lambda^2 k^2/2 +A (\hat{\varphi}_{\gamma,{\rm cri}}(k) -\hat{\varphi}_{\gamma,{\rm cri}} (0) )) -1} \; ,
\ee
while the corresponding critical chemical potential $\mu_{\gamma,{\rm cri}}$ is simply given by the ODLRO condition~(\ref{ODLROcri})
specified to the critical point.  

In presence of a condensate, the ODLRO condition~(\ref{ODLROcondition}), which can be rewritten in terms of the dimensionless 
quantities as
\be
\la{ODLRO}
\mu= A \rho + A \hat{\varphi}_\gamma (0) \; ,
\ee
is still fulfilled. The coupled equations~(\ref{PhiCond}) and (\ref{DensCond}) then become
\begin{multline}
\label{PotBis}
\hat{\varphi}_\gamma(q)= \rho_{\rm cond} \exp(-q^2/(2\gamma^2)) \\
+ \frac{\gamma^2}{\sqrt{2\pi}} \int_0^\infty \dd k \; \frac{k}{q} \frac{
\left[ \exp(-(q-k)^2/(2\gamma^2)) - \exp(-(q+k)^2/(2\gamma^2))   \right]} 
{\exp(\lambda^2 k^2/2 +A (\hat{\varphi}_\gamma(k) -\hat{\varphi}_\gamma (0) )) -1} \; 
\end{multline}
and 
\be
\la{DensBis}
\rho = \rho_{\rm cond} + \sqrt{ \frac{2}{\pi}}  \int_0^\infty \dd k \;  k^2 
\frac{1}{\exp(\lambda^2 k^2/2 +A (\hat{\varphi}_\gamma(k) -\hat{\varphi}_\gamma(0) )) -1} \; ,
\ee
where $\rho_{\rm cond}$  is now the dimensionless density of the condensate. 
One can first fix $\rho_{\rm cond} > 0$ and solve the integral equation~(\ref{PotBis}) 
which gives the effective potential $\hat{\varphi}_\gamma(q)$, and 
we then compute successively the total density
\be
\la{DensBis}
\rho = \rho_{\rm cond} + \sqrt{ \frac{2}{\pi}}  \int_0^\infty \dd k \;  k^2 
\frac{1}{\exp(\lambda^2 k^2/2 +A (\hat{\varphi}_\gamma(k) -\hat{\varphi}_\gamma(0) )) -1} \; ,
\ee
and the chemical potential by applying the ODLRO condition~(\ref{ODLRO}). 
Note that the density of the normal fluid, \textit{i.e.} $(\rho - \rho_{\rm cond})$,
does not reduce to the critical density $\rho_{\gamma,{\rm cri}}$ for  finite values of $\gamma$, in contradistinction to the cases 
$\gamma=0$ or $\gamma=\infty$. 

\subsection{Isotherms for various potential-ranges }
\label{Isotherms}

\begin{figure}
\centering
\includegraphics[width=0.8\textwidth]{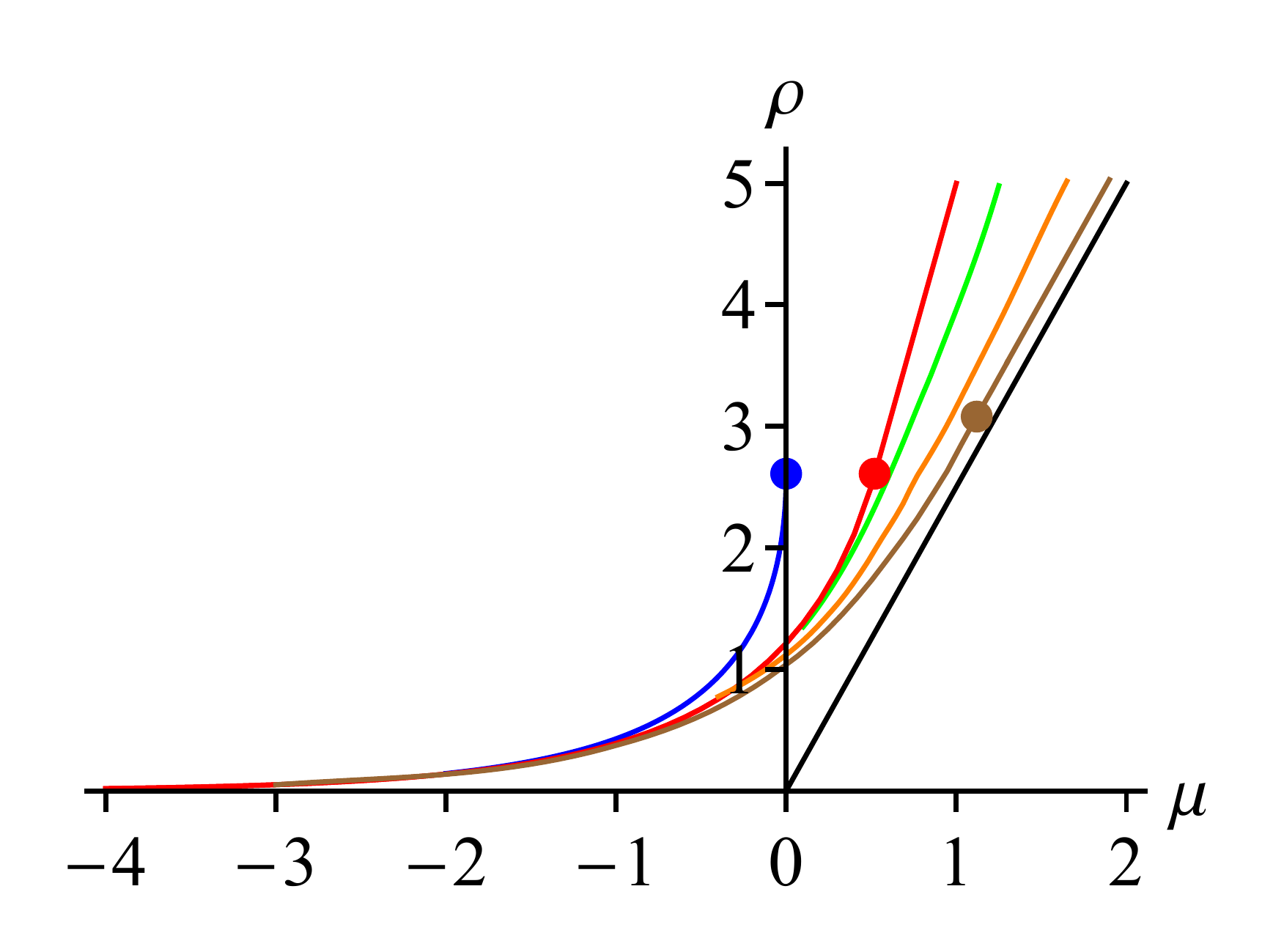}
\caption{Isotherms obtained from the numerical solution of the Hartree-Fock equations 
for a Gaussian potential characterized by  $A = 0.2$, $\lambda = 1$ and $\gamma=0.4$ (green), 
$\gamma=\gamma_0 \approx 1.0$ (orange) and $\gamma=1.9$ (brown). Additionally, the 
mean-field isotherm ($\gamma=0$) is plotted (red) alongside with the isotherm of an ideal gas (blue). The black straight line 
is defined by the equation $\mu=2 A \rho$ satisfied by
the $\gamma=\infty$ isotherm above the critical density. The critical points are marked by dots in the corresponding color.}
\label{fig:iso}
\end{figure}

We have solved the above-formulated integral equations numerically, 
using a standard Neumann method with successive overrelaxation~\cite{Kleinman1991}, 
i.e. constructing the series $\hat{\varphi}_{\gamma}^{(n)}$ with 
\begin{equation}
\hat{\varphi}_{\gamma}^{(n+1)} =  \alpha {\cal L} \hat{\varphi}_{\gamma}^{(n)} + (1-\alpha) \hat{\varphi}_{\gamma}^{(n)} 
\label{l1}
\end{equation}
where ${\cal L}$ is the integral operator in the right hand side of integral equations~\eqref{Pot},~\eqref{PotCri} or \eqref{PotBis} 
depending on whether we are below, at, or above the critical point. 
As the starting point we take the Gaussian functions, $\hat{\varphi}_{\gamma}^{(0)} = B e^{- \kappa^2 k^2}$, 
varying $B>0$ and $\kappa$ to confirm that different $\hat{\varphi}_{\gamma}^{(0)}$ converge to the same point.
The relaxation parameter, $\alpha$ was taken to be $\alpha=1/32$. 
The isotherms calculated for $A=0.2$ and $\lambda=1.0$ are shown in Fig.~\ref{fig:iso}.

\begin{figure}
\centering
\includegraphics[width=0.8\textwidth]{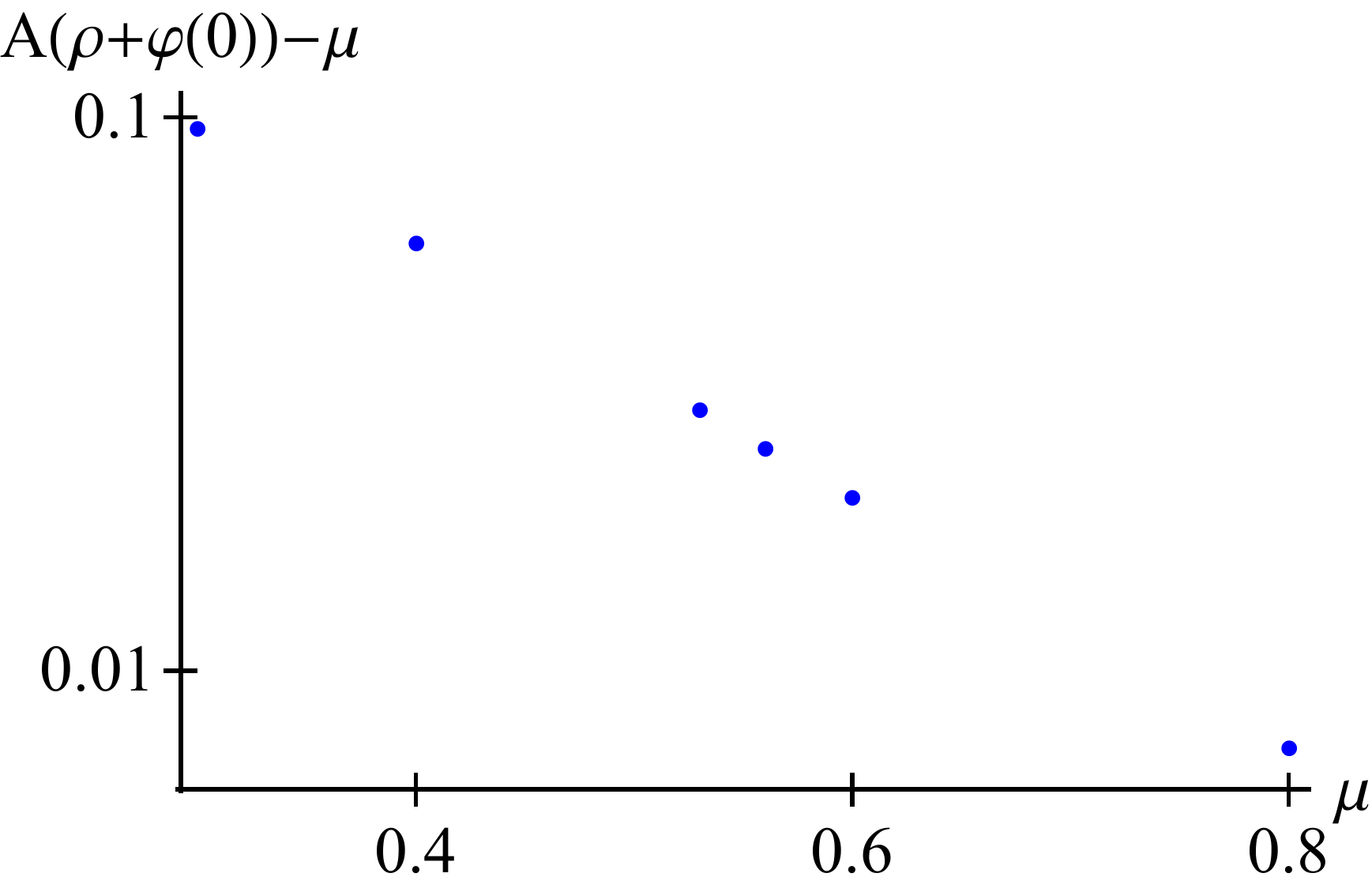}
\caption{The ODLRO parameter $A \rho + A \varphi(0) - \mu$ as a function of $\mu$ for $A=0.2$, $\lambda=1$ and $\gamma=0.4$.}
\label{fig:ODLRO}
\end{figure}

The following important 
feature emerges from those results. As $\gamma$ is reduced, 
we observe the critical point to move progressively towards larger densities up to the threshold $\gamma_0$, 
where the critical point disappears. This is manifested by the lack of convergence of the equations~\eqref{PotCri}-\eqref{DensCri}, 
and by the fact that as we move along the isotherm, the ODLRO parameter decreases, but never vanishes. 
As illustrated in Fig.~\ref{fig:ODLRO} this decrease is approximately exponential with $\mu$. 
The disappearance of the critical behavior at small $\gamma$ can, of course, only be supported, not proven, 
by numerical methods. In particular, one cannot rule out that the fixed point of Eq.~\eqref{l1} has a relatively small basin of attraction, and thus can only be reached starting from a very specific initial function $\hat{\varphi}_{\gamma}^{(0)}$. Note that for $\gamma > \gamma_0$, in 
the regions where a condensate is present, the isotherms indeed lie in the wedge defined by the straight lines 
$\rho=\mu/A$ and $\rho=\mu/(2A)$, as predicted in Section~\ref{HFapproximation} when
$\hat{v}(\bk)$ is positive for any $k$.

\subsection{Dependence of the threshold value $\gamma_0$  on the temperature}
\label{CrossoverT}

\begin{figure}
\centering
\includegraphics[width=0.8\textwidth]{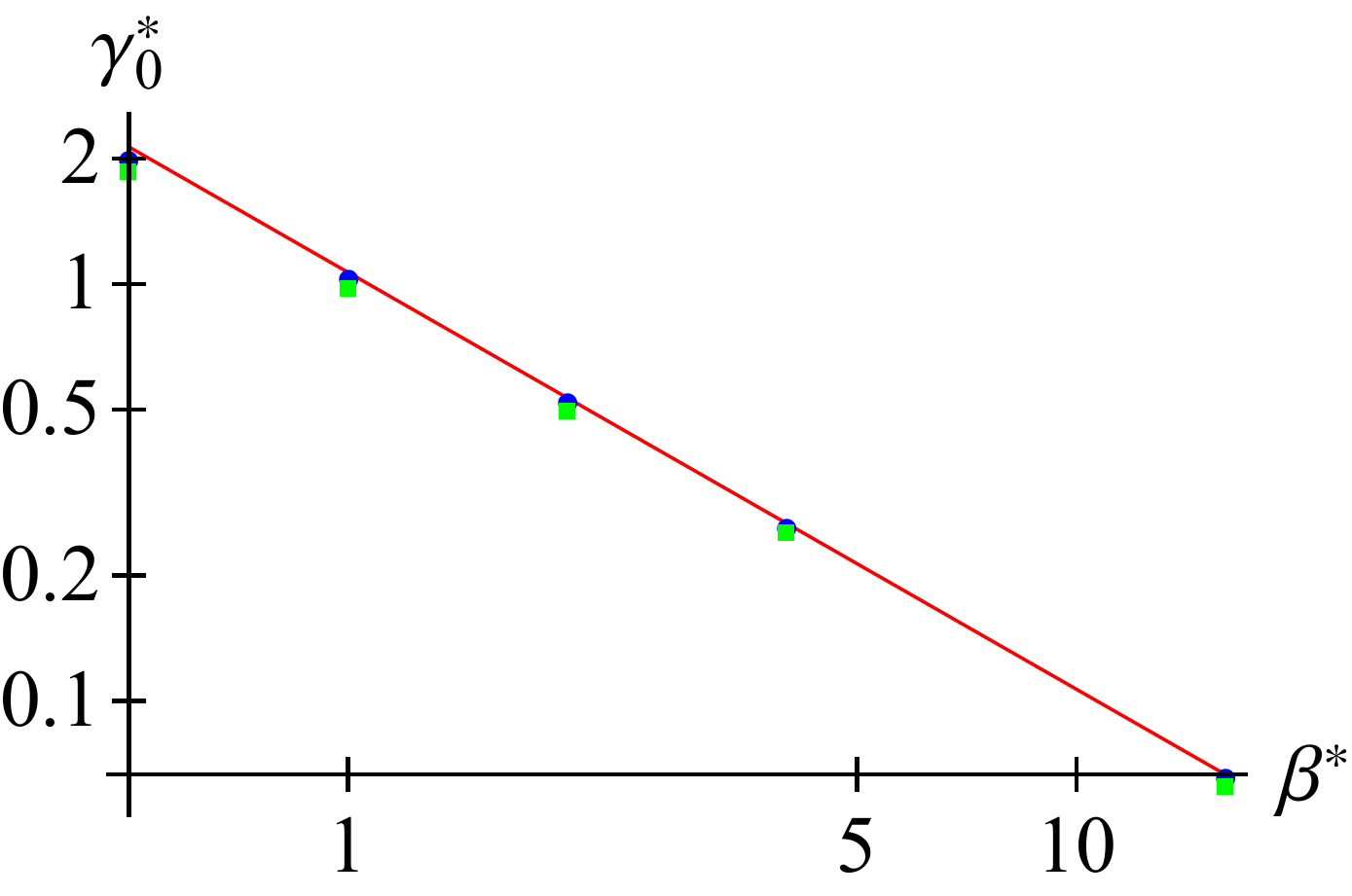}
\caption{The threshold value of $\gamma$ below which the condensation disappears as a function of the dimensionless 
temperature $\beta^*=T_{\rm ref}/T$, where $T_{\rm ref}$ is the temperature at which $A(T_{\rm ref})=0.2$ and 
$\lambda(T_{\rm ref})=1$. Green squares denote the results of the iterative solution of integral equation~\eqref{PotCri}, 
whereas blue dots are the values $\gamma_0^*$ obtained within the simplified effective-mass 
version of the Hartree-Fock equations 
presented in Appendix~\ref{EffectiveMass}. The red line shows the results of the small-T leading term~\eqref{COLT} of $\gamma_0^*(T)$
computed within that simplified version.}
\label{fig:gamma}
\end{figure}

As exposed in Section~\ref{AbsenceODLRO}, for a given potential with fixed amplitude and range, 
$\gamma_0$ is a function of $T$, $\gamma_0(T)$. 
The numerical results displayed above were obtained for some reference temperature $T_{\rm ref}$ such that 
$A=A_{\rm ref}=0.2$ and $\lambda=\lambda_{\rm ref}=1$. When the temperature is varied, the corresponding values 
of $A$ and $\lambda$ become $A=A_{\rm ref} T_{\rm ref}/T$ and $\lambda=\lambda_{\rm ref} (T_{\rm ref}/T)^{1/2}$. 
Accordingly, we compute the corresponding value of $\gamma_0(T)$ 
within the numerical method described in Section~\ref{Isotherms}, for several values of the ratio $\beta^*=T_{\rm ref}/T$. 
The results are shown in Figure~\ref{fig:gamma}, where we also plot the function $\gamma_0^*(\beta^*)$ inferred from 
the simplified effective-mass version of the Hartree-Fock equations as detailed in Appendix~\ref{EffectiveMass}. 
The effective-mass approach turns to be in good agreement with the numerical solution of the full Hartree-Fock equations. 

\begin{figure}
\centering
\includegraphics[width=0.8\textwidth]{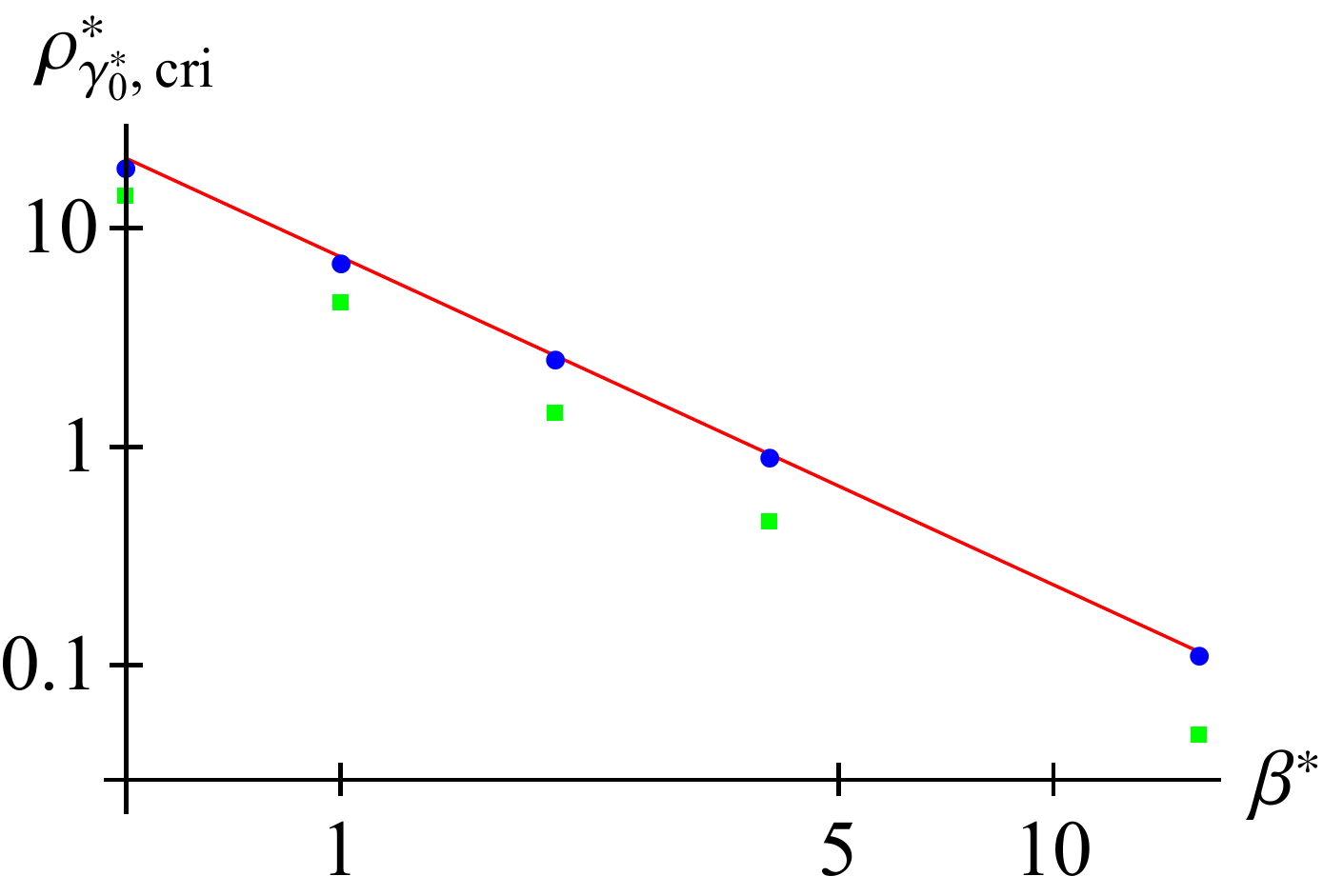}
\caption{The critical densities $\rho_{\gamma_0,{\rm cri}}$ corresponding to the threshold values of $\gamma$ 
presented in Fig.~\ref{fig:gamma}. Green squares denote the results of the iterative solution of Eq.~\eqref{PotCri}, whereas blue dots are computed  
from formula~(\ref{DensCriEM}) for $\rho_{\gamma_0^*,{\rm cri}}^*$ derived within the simplified effective-mass version of the Hartree-Fock equations.  
The red line shows the leading low-T term of $\rho_{\gamma_0^*,{\rm cri}}^*$ obtained 
by combining formula~(\ref{DensCriEM}) with the asymptotic behavior~\eqref{COLT}.}
\label{fig:rho}
\end{figure}

The computed values of $\gamma_0(T)$ indeed increase with the temperature. Moreover the 
approximate value $\gamma_0^*(T)$ increases from $0$ to $\infty$ when $T$ varies from 
$0$ to $\infty$ as shown in Appendix~\ref{EffectiveMass}. These results are consistent with the prediction 
introduced in Section~\ref{AbsenceBE}, 
namely $\gamma_0(T)$ is an increasing unbounded function of $T$. 
Thus, for a given potential with fixed amplitude and fixed range, let us say $\gamma=1$, 
there should always exist a critical temperature $T_{\rm c}$ defined by the 
equation $\gamma_0(T_{\rm c})=1$. For $T < T_{\rm c}$, the ODLRO emerges 
above some critical density $\rho_{\rm cri}(T)$, while for $T > T_{\rm c}$ the ODLRO is fully destroyed at any density. 
At $T=T_{\rm c}$, the ODLRO takes place above a critical density which is finite, and close to the effective-mass formula~(\ref{DensCriEM}) with 
$\gamma_0^*(T_{\rm c})=1$. The critical densities as a function of $\beta^*$ are plotted in Figure~\ref{fig:rho}. 
As observed, they are also increasing functions of the temperature. 

\section{Concluding remarks and comments}
\label{Conclusion}

In this paper we have computed the small-$\gamma$ leading corrections to the EOS of the mean-field model
within the HF approximation. This asymptotic analysis shows that the ODLRO cannot occur in presence of
sufficiently long-range two-body interactions. Such findings are confirmed by a numerical solution of the HF equations on the one hand, 
and by analytical calculations within a simplified effective-mass version of that approximation on the other hand. 
Note that a complete proof of the corresponding breakdown of the BE condensation for 
$\gamma$ sufficiently small within the HF approximation 
is far beyond our scope and it remains a challenging task at the mathematical level.
However, we provided here strong arguments in favor of this most interesting possibility.

An interesting byproduct of our analysis 
is the emergence of a critical temperature for a system with a given potential $v(r)$ of pair interactions. 
This prediction follows from the 
existence of the threshold $\gamma_0$ for the class of potentials $V_\gamma(r)=\gamma^3v(\gamma r)$ associated 
with the potential $v(r)$ as explained in Section~(\ref{CrossoverT}). The 
corresponding critical temperature above which the BE condensation disappears for any density 
depends both on the potential range and its amplitude. Unsurprisingly, the critical temperature increases if the potential range 
and/or its amplitude decreases.

The status of our predictions beyond the HF approximation is of course questionable. In fact, we notice that, even in 
the normal phase for $\mu < \mu_{0,{\rm cri}}$, the exact small-$\gamma$ leading corrections to the EOS of the mean-field model 
are not entirely given by the HF theory. Indeed, the correlations (neglected by the HF approximation) provide corrections 
of order $\gamma^3$ like the HF approximation itself. They have been 
first computed \textit{via} summations of Mayer graphs for the equivalent classical polymer gas~\cite{MP2003,MP2005}, 
and they were recovered within the hierarchy for the imaginary-time Green functions in Ref.~\cite{AP2011}. 
Therefore, we expect that such correlations also contribute 
to the exact small-$\gamma$ corrections near or above the mean-field critical chemical potential $\mu_{0,{\rm cri}}$, 
especially since the HF theory, which is a mean-field approximation, is unable to correctly describe the 
critical fluctuations in the vicinity of the critical point. 
However, in order to restore the BE condensation itself, there should exist an unlikely conspiracy 
induced by correlations which counteracts the interplay between long-range interactions and 
critical tails, highlighted here within the HF theory~\footnote{The interplay between quantum statistics 
and long range interactions is well illustrated through the thermodynamic equivalence between two-dimensional mean-field models 
describing fermions with attractive interactions on the one hand, and bosons with repulsive interactions on the other hand~\cite{NP2017}. }. Note that, 
beyond the predictions which might be inferred from asymptotic calculations of the exact leading small-$\gamma$ corrections, 
it would remain to perform a rigorous analysis for small but finite values 
of $\gamma$~\cite{M2005}~\footnote{The sole available mathematical result concerns the limit $\gamma \to 0$, 
for which it has been be proved that the EOS becomes 
identical to that of the mean-field model~\cite{L1986}. This is reminiscent of the derivation of the 
van der Waals equation of state for a classical fluid~\cite{KUH1963,LP1966} and its quantum version~\cite{L1966}.}. 

For very short-range potentials with $\gamma$ sufficiently large, the HF theory predicts the persistence 
of the BE condensation. No exact results are available at 
$\gamma = \infty$ and finite temperature where correlations cannot be neglected. 
Nevertheless, in this case numerical simulations~\cite{NumericalBEinfty} as well as 
theoretical calculations~\cite{TheoreticalBEinfty} strongly suggest the persistence of the BE condensation. Moreover, they provide 
systematic corrections to the critical HF quantities at low densities~\cite{PS2003}. Hence, it is instructive to notice 
that the HF theory correctly predicts the BE condensation although it fails to 
describe the exact behavior of the quantities of interest. So, if the whole
HF picture for $0 \leq \gamma \leq \infty$ is not drastically affected by correlations, a critical temperature should truly emerge. 
Interestingly, this would mimic the behavior of 
Helium IV, keeping in mind that the link between superfluidity and BE condensation is still a matter of debate.

Eventually, let us recall that our HF analysis is restricted to finite temperatures, so it does not give access to 
the properties of the ground state. We mention that 
various important  mathematical results at zero temperature 
are reviewed in the book~\cite{LSSY2005}. If most of them 
concern both cases with or without a trapping potential, 
the considered particle interactions are always short-range. 

\appendix{}
\section{Leading terms in the small-$\gamma$ expansions}
\la{ASEXP}

\subsubsection{Below the critical chemical potential for the mean-field model}
\la{asBC}
For $\mu < \mu_{0,{\rm cri}}$,
in the limit $\gamma \to 0$, $\eta_{\gamma}(\mu,\beta)$
tends to the finite value $\eta_{0}(\mu,\beta)=a\rho_{0}(\mu,\beta)-\mu > 0$ obtained for the mean-field model. Then to leading order 
$\hat{\phi}_\gamma(\bq)$ is simply given by replacing $n_\gamma^{\rm B}(\bk)$ by $n_0^{\rm B}(\bk)$ in the r.h.s. of equation~(\ref{HFas}) 
\begin{equation}
\la{PotBC}
\hat{\phi}_\gamma(\bq) \sim \frac{\gamma^3 v(0) }
{{\exp[\beta (\epsilon(\mathbf{q}) + \eta_{0}(\mu,\beta))]-1 } } \; 
\end{equation}
where the mean-field density is the solution of 
\begin{equation}
\la{MFBC}
\rho_{0}(\mu,\beta)=\frac{1}{(2\pi)^3} \int \dd \mathbf{k} 
\frac{1}{\exp[\beta (\epsilon(\mathbf{k})-\mu + a\rho_{0}(\mu,\beta))]-1 } \; .
\end{equation}
Thus $\hat{\phi}_\gamma(\bq)$ is of order $\gamma^3$. A simple algebraic equation for the leading 
term in the difference $\rho_{\gamma}(\mu,\beta)-\rho_{0}(\mu,\beta)$ 
is then readily obtained by expanding  $n_\gamma^{\rm B}(\bk)$ around $n_0^{\rm B}(\bk)$ in powers of $\beta a (\rho_{\gamma}(\mu,\beta)-\rho_{0}(\mu,\beta))$ 
and $\beta\hat{\phi}_\gamma(\bq)$ in the r.h.s. of equation~(\ref{HFdensity}). Using the leading behavior~(\ref{PotBC}), we find that such difference 
is also of order $\gamma^3$, namely 
\begin{multline}
\la{DenBC}
\rho_{\gamma}(\mu,\beta)-\rho_{0}(\mu,\beta) \sim - \frac{\gamma^3 \beta v(0)}{(2\pi)^3 [1+a\chi^{({\rm id})}(\mu - a \rho_{0}(\mu,\beta),\beta)] } \\
\int \dd \mathbf{k} 
\frac{\exp[\beta (\epsilon(\mathbf{k}) + \eta_{0}(\mu,\beta))]}{[\exp[\beta (\epsilon(\mathbf{k})+\eta_{0}(\mu,\beta))]-1]^3 } \; ,
\end{multline}
with 
\be
\la{drhodmu}
\chi^{({\rm id})}(\mu,\beta) = \frac{\partial \rho^{({\rm id})}}{\partial \mu}
\ee
where $\rho^{({\rm id})}(\mu,\beta)$ is the ideal gas density.

\subsubsection{Near the critical mean-field chemical potential}
\la{asNC}

For $\mu= \mu_{0,{\rm cri}} + \nu$ and $\rho_{\gamma}(\mu,\beta)=\rho_{\rm cri}^{(\rm id)} + n_{\gamma}(\nu,\beta)$, the definition~(\ref{eta}) 
can be then rewritten as 
\be
\la{etabis}
\eta_{\gamma}(\nu,\beta)= an_{\gamma}(\nu,\beta) -\nu \; .
\ee
Here we study the leading terms in the HF quantities for both $\nu$ and $\gamma$ small. The corresponding 
density $\rho_{\gamma}(\mu_{0,{\rm cri}} + \nu,\beta)$ is close to $\rho_{0}(\mu_{0,{\rm cri}},\beta)=\rho_{\rm cri}^{(\rm id)}$, 
so the deviation $n_{\gamma}(\nu,\beta)$ is small, as well as $\eta_{\gamma}(\nu,\beta)$. 
For small values of $\bq$ the effective energy
\be
\la{Ubis}
U_{\gamma}(\bq)= \epsilon(\mathbf{q})+ \eta_{\gamma}(\nu,\beta) + \hat{\phi}_\gamma(\mathbf{q}) \; ,
\ee
is also small, so the effective 
Bose distribution~(\ref{BEdistribution}) behaves as 
\be
\la{asBEdistribution}
n_\gamma^{\rm B}(\bq) \sim \frac{1}
{\beta U_\gamma(\bq) } \; . 
\ee 
Hence, the local equation~(\ref{HFas}) becomes a simple second-order polynomial equation 
for the leading term in $\hat{\phi}_\gamma(\mathbf{q})$ yielding the formula
\be
\label{PotNCsmall}
\hat{\phi}_\gamma(q) \sim \frac{1}{2} \left[ \sqrt{[\epsilon(q) + \eta_{\gamma}(\nu,\beta)]^2 + 4 \gamma^3 \beta^{-1} v(0) }  
-\epsilon(q) - \eta_{\gamma}(\nu,\beta)  \right] \; ,
\ee 
where $\eta_{\gamma}(\nu,\beta)$ is given by equation~(\ref{etabis}).
For finite values of $q$ 
the kinetic energy $\epsilon(\mathbf{q})$ dominates all other contributions in $U_{\gamma}(\bq)$, and the 
leading term in $\hat{\phi}_\gamma(q) $ takes the form similar to that given by~(\ref{PotBC}), \textit{i.e.} 
 \footnote{Interestingly, the behavior~(\ref{PotNCfinite}) also holds for small values of $q$ such that $\epsilon(\mathbf{q})$ nevertheless dominate 
the contributions of $\hat{\phi}_\gamma(\mathbf{q})$ and of $\eta_{\gamma}(\nu,\beta)$ in $U_{\gamma}(\bq)$. Then this 
behavior becomes identical to that of the small-$q$ 
expression~(\ref{PotNCsmall}). Hence formulae~(\ref{PotNCsmall}) and 
~(\ref{PotNCfinite}) perfectly match at leading order for small intermediate values of $q$, and 
they provide a complete description of $\hat{\phi}_\gamma(\mathbf{q})$ 
in the whole range $0 \leq q < \infty$.}
\begin{equation}
\la{PotNCfinite}
\hat{\phi}_\gamma(\bq) \sim \frac{\gamma^3 v(0) }
{{\exp[\beta \epsilon(\mathbf{q})]-1 } } \; .
\end{equation}

Once the leading behavior of $\hat{\phi}_\gamma(\mathbf{q})$ has been determined, it remains to determine self-consistently the deviation
$n_{\gamma}(\nu,\beta)$. For this purpose we first rewrite $n_{\gamma}(\nu,\beta)$ as
\be
\la{DensDevNC}
n_{\gamma}(\nu,\beta) =  \frac{1}{2\pi^2} \int_0^\infty \dd q \;  q^2 
\frac{\exp(\beta \epsilon(q))\left[1- \exp(\beta (\hat{\phi}_{\gamma}(q) + \eta_{\gamma}(\nu,\beta))\right]}
{(\exp(U_{\gamma}(q)) -1)(\exp(\beta \epsilon(q)) -1)} \; ,
\ee
where we have used the formula
\be
\la{DensIdC}
\rho_{\rm cri}^{(\rm id)} =  \frac{1}{2\pi^2} \int_0^\infty \dd q \;  q^2 
\frac{1}
{[\exp(\beta \epsilon(q)) -1]} \; 
\ee
for the ideal gas critical density. In the integral~(\ref{DensDevNC}), the leading contributions are expected to arise from 
the neighbourhood  of $q=0$.  Accordingly, the behavior of $n_{\gamma}(\nu,\beta)$ to leading order reduces to 
\be
\la{DensDevNCleading}
n_{\gamma}(\nu,\beta) \sim  -\frac{m}{\pi^2 \beta \hbar^2} \int_0^{ \infty } \dd q \;   
\frac{(\hat{\phi}_{\gamma}(q) + \eta_{\gamma}(\nu,\beta))}{U_{\gamma}(q)}\;  
\ee
which is obtained by expanding all Boltzmann factors involved in the integral~(\ref{DensDevNC}) in powers of the small quantities 
$\epsilon(\mathbf{q})$, $\eta_{\gamma}(\nu,\beta)$, $\hat{\phi}_\gamma(\mathbf{q})$ and $U_{\gamma}(q)$. 
Note that the integral~(\ref{DensDevNCleading}) does converge when $q \to \infty$ thanks to 
the presence of the kinetic energy $\epsilon(q)$ in $U_{\gamma}(q)$. Inserting the expression~(\ref{PotNCsmall}) for 
$\hat{\phi}_{\gamma}(q)$ into the integral~(\ref{DensDevNCleading}), and making
the change of variable 
$q \rightarrow \theta$ defined by 
\begin{equation}
\label{VarCha}
\epsilon(q) + \eta_{\gamma}(\nu,\beta) = 2 \sqrt{\beta^{-1} v(0)} \gamma^{3/2} \sinh \theta \; ,
\end{equation}
we find 
\begin{equation}
\label{PotNCsmalltheta}
\hat{\varphi}_\gamma(\theta) = \sqrt{\beta^{-1} v(0)} \gamma^{3/2} e^{-\theta}   
\end{equation}
and 
\begin{equation}
\label{Ubistheta}
U_\gamma(\theta) = \sqrt{\beta^{-1} v(0)} \gamma^{3/2} e^{\theta}   \; ,
\end{equation}
so the consistency equation for $n_{\gamma}(\nu,\beta)$ becomes to leading order                                                                     
\be
\la{DensDevNCleadingbis}
n_{\gamma}(\nu,\beta) \sim  \frac{1}{\pi^2\lambda_{\rm dB}^3} [\beta v(0)]^{1/4} \gamma^{3/4} K(\theta_0)
\ee
with de Broglie thermal wavelength $\lambda_{\rm dB} = (\beta \hbar^2/m)^{1/2}$ and the dimensionless function $K(\theta_0)$ defined by the convergent integral
\begin{equation}
\label{Kfunction}
K(\theta_0) = - \int_{\theta_0}^{\infty} \dd \theta \; \frac{\cosh \theta}{(\sinh \theta -\sinh \theta_0)^{1/2}} 
(e^{-2\theta} + 2 e^{-\theta} \sinh \theta_0 )\; ,
\end{equation}
while $\theta_0$ is such that
\begin{equation}
\label{thetamin}
\eta_{\gamma}(\nu,\beta) = a n_{\gamma}(\nu,\beta) -\nu = 2 \sqrt{\beta^{-1} v(0)} \gamma^{3/2} \sinh \theta_0 \; .
\end{equation}
The function $K(\theta_0)$ is a monotonously decaying function which varies from $\infty$ to $-\infty$ when $\theta_0$
varies from $-\infty$ to $\infty$. Hence its inverse $K^{-1}$ exists and is well defined. Moreover, $K(\theta_0)$ behaves as 
\begin{equation}
\label{Kasy}
K(\theta_0) \sim \frac{2\sqrt{2}}{15} e^{-5\theta_0/2} 
\end{equation}
when $\theta_0 \to -\infty$.

In order to solve self-consistently equation~(\ref{DensDevNCleadingbis}) for $n_{\gamma}(\nu,\beta)$,  
it is useful to scale $\nu$ as a function of $\gamma$, typically as a power law. According to the presence of the multiplicative factor 
$\gamma^{3/4}$ in the consistency equation~(\ref{DensDevNCleadingbis}), and since $K(\theta_0)$ does not depend explicitly on $\gamma$, it is convenient to
scale $\nu$ as $\nu = \nu^\ast \gamma^{3/4}$. Within this scaling, we look for the power law
behavior $n_{\gamma}(\nu,\beta) \sim n^\ast \gamma^{s}$ with $s > 0$:

\begin{itemize}

\item If $s$ was smaller than $3/4$, $\eta_{\gamma}(\nu,\beta)$ 
would behave as $a n^\ast \gamma^{s} $, so $\theta_0$ would diverge as 
${\rm cst} \gamma^{s-3/2}$ with the same sign as $n^\ast$: each side of the consistency equation~(\ref{DensDevNCleadingbis}) would then have opposite signs, 
so the powers $s < 3/4$ are excluded. 

\item If $s$ was larger than $3/4$, $\eta_{\gamma}(\nu,\beta)$ would behave as $-\nu^\ast \gamma^{3/4} $, so 
$\theta_0$ would diverge as ${\rm cst} \gamma^{-3/4}$: the r.h.s. of the consistency equation~(\ref{DensDevNCleadingbis}) would then become much larger than
the l.h.s., so the powers $s > 3/4$ are excluded. 

\end{itemize}

Hence, the sole possible power law is $s=3/4$. Then, the consistency equation~(\ref{DensDevNCleadingbis}) 
becomes,
\be
\la{ThetaNu}
\nu^\ast =  \frac{a}{\pi^2\lambda_{\rm dB}^3} [\beta v(0)]^{1/4}  K(\theta_0^{(0)}) \; . 
\ee
with the finite $\theta_0^{(0)} = \lim_{\gamma \to 0} \theta_0$. Since the inverse $K^{-1}$ exists, equation~(\ref{ThetaNu})
has the unique solution $\theta_0^{(0)}=K^{-1}\left( \pi^2\lambda_{\rm dB}^3 [\beta v(0)]^{-1/4} \nu^\ast /a  \right) $.
Note that $\theta_0^{(0)}$ is of order $\gamma^0$, so $\eta_{\gamma}(\nu,\beta) = a n_{\gamma}(\nu,\beta) -\nu$ given by formula~(\ref{thetamin}) 
is of order $\gamma^{3/2}$. Hence we see that the consistency equation~(\ref{DensDevNCleadingbis}) not only determines 
the leading term of order $\gamma^{3/4}$ in the small-$\gamma$ expansion of 
$n_{\gamma}(\nu,\beta)$ for $\nu = \nu^\ast \gamma^{3/4}$, but also the first subleading correction of order $\gamma^{3/2}$. 
This leads to formula~(\ref{DensDevNCexpansion}).

\subsubsection{Above the critical mean-field chemical potential}
\la{asAC}
 
Now we fix  $\mu > \mu_{0,{\rm cri}} $, and as above 
we still define $\nu= \mu - \mu_{0,{\rm cri}} $ and $n_{\gamma}(\nu,\beta)= \rho_{\gamma}(\nu,\beta) - \rho_{\rm cri}^{(0)} $. 
In the integral~(\ref{DensDevNC}), the leading contributions are again expected to arise from 
the neighbourhood  of $q=0$, so after making the variable change~(\ref{VarCha}), the consistency equation becomes
\be
\la{DensAC}
\frac{1}{\pi^2\lambda_{\rm dB}^3} [\beta v(0)]^{1/4} \lim_{\gamma \to 0}  \gamma^{3/4} K(\theta_0) = \nu/a \; ,
\ee
since the deviation $n_{\gamma}(\nu,\beta)$ goes to  $\nu/a$ when $\gamma \to 0$. 

The consistency equation~(\ref{DensAC}) now implies that $K(\theta_0)$ goes to $\infty$ when $\gamma \to 0$, so $\theta_0$ goes to $-\infty$ 
in this limit. From equation~(\ref{thetamin}), 
we infer that $\eta_{\gamma}(\nu,\beta)$ is negative and goes to $0$ slower than $\gamma^{3/2}$. 
Inserting into the l.h.s. of equation~(\ref{DensAC}) the 
asymptotic behavior~(\ref{Kasy}), we eventually obtain 
\begin{equation}
\label{CorrEOS}
\eta_{\gamma}(\nu,\beta) \sim - \left( \frac{15 \pi^2 \lambda_{\rm dB}^3  \nu}{2 \sqrt{2} \beta^{3/2} v(0)^{3/2} a} \right)^{2/5} v(0) \gamma^{6/5} \; .
\end{equation}
Like in the vicinity of the critical point, the consistency equation~(\ref{DensAC}) determines the first two terms in the small-$\gamma$
expansion~(\ref{DensDevACexpansion}) of $n_{\gamma}(\nu,\beta)$.

\section{Subleading terms in the small-$\gamma$ expansions}
\label{SubleadingCorrections}

\subsection{Decomposition of $n_{\gamma}(\nu,\beta)$  }

In order to separate the contributions to $n_{\gamma}(\nu,\beta)$ of small values of $q$ on the one hand, from those of finite values of $q$ 
on the other hand, it is useful to introduce 
the exact decomposition 
\be
\la{DensDevNCbis}
n_{\gamma}(\nu,\beta) = n_{\gamma}^{(1)}(\nu,\beta) +  n_{\gamma}^{(2)}(\nu,\beta)   \; ,
\ee
with
\be
\la{DensDevNCsmall}
n_{\gamma}^{(1)}(\nu,\beta) =  -\frac{m}{\pi^2 \beta \hbar^2} \int_0^{ \infty } \dd q \;   
\frac{(\hat{\phi}_{\gamma}(q) + \eta_{\gamma}(\nu,\beta))}{U_{\gamma}(q)}\;  
\ee
and
\begin{multline}
\la{DensDevNCfinite}
n_{\gamma}^{(2)}(\nu,\beta)  =  \frac{1}{2\pi^2} \int_0^\infty \dd q \; [  q^2 
\frac{\exp(\beta \epsilon(q))\left[1- \exp(\beta (\hat{\phi}_{\gamma}(q) + \eta_{\gamma}(\nu,\beta)))\right]}
{(\exp(\beta U_{\gamma}(q)) -1)(\exp(\beta \epsilon(q)) -1)} \\ + \frac{2m}{\beta \hbar^2} 
\frac{(\hat{\phi}_{\gamma}(q) + \eta_{\gamma}(\nu,\beta))}{U_{\gamma}(q)} ] \; .
\end{multline}

The leading contributions of $n_{\gamma}^{(1)}(\nu,\beta)$ have been studied in Appendix~\ref{ASEXP}. In the following we determine 
the corresponding subleading corrections, as well as the leading contribution of $n_{\gamma}^{(2)}(\nu,\beta)$.

\subsection{Near the critical point}

\subsubsection{Corrections arising from $n_{\gamma}^{(1)}(\nu,\beta)$ }

The various corrections to the previous estimation derived in Appendix~\ref{ASEXP} arise from :

\begin{itemize}

\item (a) The variation 
\be
\la{deltaeta}
\delta \eta_{\gamma}(\nu,\beta) = \eta_{\gamma}(\nu,\beta) -\eta_{\gamma}^{(0)}(\nu,\beta)
\ee
where $\eta_{\gamma}^{(0)}(\nu,\beta)$ is the leading term of order $\gamma^{3/2}$ in the small-$\gamma$ expansion of $\eta_{\gamma}(\nu,\beta)$, 
as well as the corresponding variation $\delta \theta_0$ of $\theta_0$ defined by equation~(\ref{thetamin}), 
\be
\la{deltatheta}
\delta \theta_0 = \theta_0 - \theta_0^{(0)} 
\ee
where $\theta_0^{(0)}$ is independent of $\gamma$. This induces a correction to the leading behavior of $n_{\gamma}^{(1)}(\nu,\beta)$
which behaves as 
\be
\la{DensDevNCdeltatheta}
\frac{1}{\pi^2\lambda_{\rm dB}^3} [\beta v(0)]^{1/4} K'(\theta_0^{(0)}) \gamma^{3/4} \delta \theta_0
\ee
whith $ K'(\theta) =\dd K/\dd \theta$. 

\item (b) The terms neglected in the replacement of $n_\gamma^{\rm B}(\bq)$ by its asymptotic form~(\ref{asBEdistribution}) 
into the local equation~(\ref{HFas}). They are obtained by expanding $n_\gamma^{\rm B}(\bq)$ in powers of $\beta U_\gamma(\bq)$, 
\textit{i.e.}
\be
\la{asBEdistributionBis}
n_\gamma^{\rm B}(\bq) =\frac{1}
{\beta U_\gamma(\bq) } - \frac{1}{2} +... \; . 
\ee
The corresponding  first correction $\delta \hat{\phi}_\gamma^{(b)}(q)$ to the leading potential $\hat{\phi}_{\gamma}^{(0)}(q)$ given by 
formula~(\ref{PotNCsmall}) is associated with the $1/2$-term in 
expansion~(\ref{asBEdistributionBis}). It is readily expressed in terms of $\hat{\phi}_{\gamma}^{(0)}(q)$ and $\eta_{\gamma}^{(0)}(\nu,\beta)$. 
The resulting contribution to $n_{\gamma}^{(1)}(\nu,\beta)$ then is computed 
within the variable change~(\ref{VarCha}) with $\eta_{\gamma}(\nu,\beta)$ replaced by $\eta_{\gamma}^{(0)}(\nu,\beta)$. It takes the following form 
\be
\la{Correctionsn1}
{\rm cst} \; \gamma^{9/4} K_1(\theta_0^{(0)} ) \; ,
\ee
where the function $K_1(\theta_0^{(0)} ) $ 
does not depend on $\gamma$, and it is defined as a convergent integral over $\theta$ from $\theta_0^{(0)}$ to 
$\infty$ similar to the one~(\ref{Kfunction}) defining $K(\theta_0^{(0)} ) $.

\item (c) the terms neglected in the calculation of the integral~(\ref{HFpotentialBis}) 
by keeping only the first term $n_\gamma^{\rm B}(\bq)$ in the Taylor expansion 
\be
\la{BEexpansion}
n_\gamma^{\rm B}(\bq + \gamma \bl) = n_\gamma^{\rm B}(\bq) + \gamma (\bl \cdot \nabla_{\bq}) n_\gamma^{\rm B}(\bq) + 
\frac{\gamma^2}{2} (\bl \cdot \nabla_{\bq})^2 n_\gamma^{\rm B}(\bq) +...
\ee
The first correction beyond the leading term $ n_\gamma^{\rm B}(\bq)$ 
provides a vanishing contribution once it is integrated over $\bl$ thanks to the spherical symmetry of the potential, 
\textit{i.e.} $\hat{v}(\bl)=\hat{v}(l)$. The first non-vanishing corrections to the effective potential 
$\hat{\phi}_{\gamma}(q)$ defined as the solution of equation (\ref{HFas}) are then smaller by a factor $\gamma^2/(\gamma^{3/4})^2=\gamma^{1/2}$
for small $q$'s of order $\gamma^{3/4}$ on the one hand, and by a factor $\gamma^2/(\gamma^0)^2=\gamma^{2}$ for finite $q$'s on the other hand. 
They are indeed small corrections because the leading effective potential $\hat{\phi}_{\gamma}^{(0)}(q)$ does vary on larger scales than $\gamma$ 
for any $q$. The resulting correction $\delta \hat{\phi}_\gamma^{(c)}(q)$ is expressed in terms of $\hat{\phi}_{\gamma}^{(0)}(q)$, 
$\dd \hat{\phi}_{\gamma}^{(0)}(q)/\dd q$, $\dd^2 \hat{\phi}_{\gamma}^{(0)}(q)/\dd q^2$ and $\eta_{\gamma}^{(0)}(\nu,\beta)$, 
with an amplitude proportional to the finite second moment $\int \dd \bl \hat{v}(\bl) \bl^2 $ 
of the potential $\hat{v}$ in Fourier space. The resulting contribution to $n_{\gamma}^{(1)}(\nu,\beta)$ is again computed 
within the variable change~(\ref{VarCha}) with $\eta_{\gamma}(\nu,\beta)$ replaced by $\eta_{\gamma}^{(0)}(\nu,\beta)$. It reads
\be
\la{Correctionsn2}
{\rm cst} \; \gamma^{5/4} K_2(\theta_0^{(0)} ) \; , 
\ee
where the function $K_2(\theta_0^{(0)} )$ only depends on $\theta_0^{(0)}$, and it is defined by an 
integral representation similar to that~(\ref{Kfunction}) defining $K(\theta_0^{(0)} ) $.

\end{itemize}

\subsubsection{Corrections arising from $n_{\gamma}^{(2)}(\nu,\beta)$}

In the integral~(\ref{DensDevNCfinite}) we can replace, at leading order   
$(1-\exp(\beta (\hat{\phi}_{\gamma}(q) + \eta_{\gamma}(\nu,\beta))))$ by 
$-\beta (\hat{\phi}_{\gamma}(q) + \eta_{\gamma}(\nu,\beta))$ since $(\hat{\phi}_{\gamma}(q) + \eta_{\gamma}(\nu,\beta))$
remains small for any $q$. Then, at leading order $U_\gamma(q)$ can be replaced by $\epsilon(q)$, so $n_{\gamma}^{(2)}(\nu,\beta)$ 
behaves as 
\be
\la{DensDevNCfinite1}
\frac{-\beta}{2\pi^2} \int_0^\infty \dd q \;  q^2 (\hat{\phi}_{\gamma}(q) + \eta_{\gamma}(\nu,\beta))
\left[\frac{\exp(\beta \epsilon(q))}{(\exp(\beta \epsilon(q)) -1)^2} -\frac{1}{\beta^2 \epsilon(q)^2} \right] \; .
\ee
Taking into account the expressions~(\ref{PotNCsmall}) and~(\ref{PotNCfinite}), we find that the contributions of $\hat{\phi}_{\gamma}(q)$ 
to the integral~(\ref{DensDevNCfinite1}) are of order $o(\gamma^{3/2})$. Since $\eta_{\gamma}(\nu,\beta)$ is of order 
$\gamma^{3/2}$ within the scaling $\nu = \nu^\ast \gamma^{3/4}$, the leading behavior of $n_{\gamma}^{(2)}(\nu,\beta)$ then reduces to 
\be
\la{DensDevNCfinite2}
n_{\gamma}^{(2)}(\nu,\beta) \sim \frac{-\beta \eta_{\gamma}^{(0)}(\nu,\beta)}{2\pi^2} \int_0^\infty \dd q \;  q^2 
\left[\frac{\exp(\beta \epsilon(q))}{(\exp(\beta \epsilon(q)) -1)^2} -\frac{1}{\beta^2 \epsilon(q)^2} \right] \; ,
\ee
where the remaining integral over $q$ does converge and does not depend on $\gamma$. Hence $n_{\gamma}^{(2)}(\nu,\beta)$ is of order 
$\gamma^{3/2}$.  

\subsubsection{Subleading terms of order $o(\gamma^{3/2})$ for $n_{\gamma}(\nu,\beta)$}

Eventually, the lowest order corrections to the leading term of order $\gamma^{3/4}$ in the small-$\gamma$  
expansion of $n_{\gamma}(\nu,\beta)$ are given by the sum of contributions~(\ref{DensDevNCdeltatheta}), 
(\ref{Correctionsn1}), (\ref{Correctionsn2}), and~(\ref{DensDevNCfinite2}). The consistency of the calculation imposes that 
the next correction to this $\gamma^{3/4}$-term reduces to the 
$\gamma^{3/2}$-term computed in formula~(\ref{DensDevNCexpansion}). This implies that $\delta \theta_0$ is of order $\gamma^{1/2}$,  
\be
\la{deltathetabis}
\delta \theta_0 = {\rm cst} \; \gamma^{1/2} K_2(\theta_0^{(0)})/K'(\theta_0^{(0)}) + O(\gamma^{3/4}) \; 
\ee
(note that $K'(\theta_0^{(0)})$ never vanishes), by setting that contribution~(\ref{DensDevNCdeltatheta})   
cancels the contribution~(\ref{Correctionsn2}) of order $\gamma^{5/4}$. In the expansion of  $\delta \theta_0$, 
the next term is of order $\gamma^{3/4}$, as shown  by 
setting that its contribution of order $\gamma^{3/2}$ in formula~(\ref{DensDevNCdeltatheta}), plus that of 
contribution~(\ref{DensDevNCfinite2}), does reduce to the 
$\gamma^{3/2}$-term in formula~(\ref{DensDevNCexpansion}). Hence, the next correction to this $\gamma^{3/2}$-term in the 
small-$\gamma$  expansion of $n_{\gamma}(\nu,\beta)$ is of order $\gamma^2$. It entirely arises from the 
corrections to the purely local equation~(\ref{HFas}) derived from the integral equation~(\ref{HFpotentialBis}).

\subsection{Above the critical point}

The subleading corrections of order $o(\gamma^{6/5})$ in the expansion~(\ref{DensDevACexpansion}) can be computed along similar lines 
as above. The leading contribution of $n_{\gamma}^{(2)}(\nu,\beta)$ is still given by formula
(\ref{DensDevNCfinite2}) with now $\eta_{\gamma}^{(0)}(\nu,\beta)$ of order $\gamma^{6/5}$. In order to compute the corrections due to the variation 
$\delta \theta_0$ in the expression of $n_{\gamma}^{(1)}(\nu,\beta)$, we now 
have to use the asymptotic behavior~(\ref{Kasy}) of $K(\theta_0)$, including 
subleading corrections proportional to  $e^{-\theta_0/2}$, since $\theta_0 \to -\infty$. Expression~(\ref{DensDevNCdeltatheta}) is then replaced here 
by the sum of two terms, namely
\be
\la{DensDevNCdeltatheta1} 
{\rm cst} \; \gamma^{-6/5} \delta \eta_{\gamma}(\nu,\beta)
\ee
which arises from the variation of the leading term proportional to $e^{-5\theta_0/2}$ in $K(\theta_0)$,  plus the 
contribution of the subleading term proportional to  $e^{-\theta_0/2}$,
\be
\la{DensDevNCdeltatheta2}
{\rm cst} \; \gamma^{3/5} \; ,
\ee
computed with $\theta_0^{(0)}$ in place of  $\theta_0$. The contributions to $n_{\gamma}^{(1)}(\nu,\beta)$ 
of corrections (b) and (c) are still given by formulae~(\ref{Correctionsn1}) and~(\ref{Correctionsn2}). But now, 
similarly to $K(\theta_0)$, both functions $K_1(\theta_0 )$ and $K_2(\theta_0 )$ diverge when $\theta_0 \to -\infty$, and they become 
proportional to $e^{-3\theta_0/2}$. The resulting contributions behave as
\be
\la{Correctionsn1AC}
{\rm cst} \; \gamma^{9/5} 
\ee
for the corrections (b), and 
\be
\la{Correctionsn2AC}
{\rm cst} \; \gamma^{4/5} 
\ee
for the corrections (c). The variation $\delta \eta_{\gamma}(\nu,\beta)$ follows by imposing that the sum of contributions~(\ref{DensDevNCfinite2}), 
(\ref{DensDevNCdeltatheta1}), (\ref{DensDevNCdeltatheta2}),
(\ref{Correctionsn1AC}) and~(\ref{Correctionsn2AC}), reduces to the first correction of order $\gamma^{6/5}$ in the small-$\gamma$ 
expansion~(\ref{DensDevACexpansion}) of $n_{\gamma}(\nu,\beta)$. The next correction in this expansion is then found to be of order 
$\gamma^{9/5}$, and is such that the contribution~(\ref{DensDevNCdeltatheta2}) is canceled by the leading behavior of 
the contribution~(\ref{DensDevNCdeltatheta1}). Note that the leading effective potential $\hat{\phi}_{\gamma}^{(0)}(q)$ 
still varies on larger scales than $\gamma$, so the introduction of the purely local equation~(\ref{HFas}) is indeed 
justified at leading order. In fact the corresponding  
corrections in the small-$\gamma$ expansion of $n_{\gamma}(\nu,\beta)$ are of order $\gamma^2$, and 
they follow by canceling the sum of contributions~(\ref{Correctionsn2AC}) and~(\ref{DensDevNCdeltatheta1}).

\section{Simplified analysis of critical properties within an effective-mass approach}
\label{EffectiveMass}

In this appendix  we proceed to an estimation, within a 
simplified version of the Hartree-Fock equations, of the threshold $\gamma_0$ for the Gaussian potential studied 
in Section~\ref{NR}. Such a simplified version of HF 
relies on a very common approximation in condensed matter which takes the full account of 
interactions through an effective mass. Here, it involves expanding the 
effective potential $\hat{\varphi}_\gamma(k)$ in powers of $k$ and keeping only 
the $k^2$-term. Denoting $\hat{\varphi}_\gamma''(0)= \partial^2 \hat{\varphi}_\gamma/\partial q^2(0)$, 
the BE distribution~(\ref{BE}) is then replaced by 
\be
\la{BEeffectivemass}
n_\gamma^{\ast}(k)=\frac{1}{\exp((\lambda^2 + A \hat{\varphi}_\gamma''(0))k^2/2 +A \hat{\varphi}_\gamma(0) +A \rho -\mu) -1} \; ,
\ee
\bigskip
which can be interpreted as describing ideal bosons in a constant potential, while their effective mass $m^*$ 
is such that the 
corresponding dimensionless de Broglie wavelength reduces to 
\be
\la{dBeff}
\lambda^* =(\lambda^2 + A \hat{\varphi}_\gamma''(0))^{1/2} \; .
\ee
At the critical point, if it exists, the ODLRO condition~(\ref{ODLROcri}) is satisfied, and the resulting critical 
BE distribution~(\ref{BEeffectivemass}) reduces to
\be 
\la{BEemCri}
n_{\gamma,{\rm cri}}^*(k)=\frac{1}{\exp((\lambda^2 + A \hat{\varphi}_{\gamma,{\rm cri}}''(0))k^2/2) -1} \; .
\ee
If we insert this BE distribution into the exact expression of $\hat{\varphi}_\gamma''(0)$, obtained by taking the second partial derivative 
with respect to $q$ of the integral expression~(\ref{PotCri}), we find the implicit equation 
\be
 \label{ImplicitCurv}
 \hat{\varphi}_{\gamma,{\rm cri}}''(0) = \gamma \sqrt{\frac{2}{\pi}}\int_0^\infty dk\, k^2 
 [-1 + k^2/3]\, \frac{\exp(-k^2/2)}{\exp(\gamma^2(\lambda^2 + A \hat{\varphi}_{\gamma,{\rm cri}}''(0))k^2/2) -1} \; . 
  \ee
After a standard and straightforward calculation, we can recast this equation as  
\be
 \label{ImplicitCurvBis}
 \hat{\varphi}_{\gamma,{\rm cri}}''(0) = \frac{2\gamma \xi}{3} \frac{\dd L}{\dd \xi}  
  \ee
with $\xi=\gamma^2(\lambda^2 + A \hat{\varphi}_{\gamma,{\rm cri}}''(0))$ and the function
\be
\la{functionL}
L(\xi) = \sum_{n=0}^\infty \frac{1}{(1 + (n+1)\xi)^{3/2}} \; , 
\ee
or equivalently as 
\be
\la{ImplicitXi}
\xi-\frac{2A\gamma^3}{3} \xi \frac{\dd L}{\dd \xi} = \gamma^2 \lambda^2 \; . 
\ee

Using the asymptotic behaviors $L(\xi) \sim 2/\xi$ when $\xi \to 0$ and 
$L(\xi) \sim \zeta(3/2)/\xi^{3/2}$ when $\xi \to \infty$, we find that the function 
\be
\la{fXi}
M(\xi)=\xi-\frac{2A\gamma^3}{3} \xi \frac{\dd L}{\dd \xi} 
\ee
decays from $\infty$ to some positive minimum $M(\xi_0)$ when $\xi$ varies from $0$ to some $\xi_0$, while it increases 
from $M(\xi_0)$ to $\infty$ when $\xi$ varies from $\xi_0$ to $\infty$. For $\gamma$ small, $M(\xi_0)$
behaves as $4 A^{1/2} \gamma^{3/2}/\sqrt{3}$, hence it is much larger than $\gamma^2 \lambda^2$. Thus, for 
$\gamma$ sufficiently small, the implicit equation~(\ref{ImplicitXi}) has no solutions, and the ODLRO 
cannot exist, in agreement with the analysis of Section~\ref{AbsenceODLRO}.

When $\gamma$ increases, a solution of equation~(\ref{ImplicitXi}) appears at a value $\gamma_0^*$ such that 
\be
\la{CrossOverEM}
\xi_0-\frac{2A\gamma_0^3}{3} \xi_0 \frac{\dd L}{\dd \xi}(\xi_0) = (\gamma_0^*)^2 \lambda^2 \; ,  
\ee
so for $\gamma \geq \gamma_0^*$, the ODLRO can exist. Note that the corresponding critical density 
is finite and reduces to 
\be 
\la{DensCriEM}
\rho_{\gamma_0^*,{\rm cri}}^*= \sqrt{ \frac{2}{\pi}}  \int_0^\infty \dd k \;  k^2 \frac{1}{\exp(\xi_0 k^2/(2 (\gamma_0^*)^2)) -1} 
=\frac{\zeta(3/2) (\gamma_0^*)^3}{\xi_0^{3/2}} \; .
\ee 
It is larger than its ideal counterpart $\rho_{\rm cri}^{(\rm id)}=\zeta(3/2)/\lambda^3$, since $(\gamma_0^*)^3/\xi_0^{3/2} > 1/\lambda^3$ 
as inferred from equation~(\ref{CrossOverEM}) and $\dd L/\dd \xi < 0$.
Like $\gamma_0$ itself, its present approximate value $\gamma_0^*$ depends on $g$ (since $A=g\lambda^3$)
and $\lambda$. Therefore $\gamma_0^*(g,\lambda)$ can be also viewed as a function of temperature, $\gamma_0^*(T)$, 
for a given Gaussian potential $v(r)$.  
At finite temperatures, $\gamma_0^*$ is computed numerically by plotting the function $M(\xi)$ (see results in Section~\ref{CrossoverT}). 
The analytical behavior of $\gamma_0^*(T)$ at low and high temperatures respectively can be determined as follows. 
 
At low temperatures, both $A=g\lambda^3$ and $\lambda$ diverge. The inspection of equation~(\ref{CrossOverEM}) 
shows then that both $\gamma_0^*(T)$ and $\xi_0$ vanish. Using the small-$\xi$ behavior of $L(\xi)$, 
we find successively $\xi_0 \sim 2 g^{1/2} \lambda^{3/2} (\gamma_0^*)^{3/2}/\sqrt{3}$ and 
\be
\la{COLT}
\gamma_0^*(T) \sim \frac{16 g}{3 \lambda}   \sim {\rm cst} \; T \;\; \text{when} \;\; T \to 0 \; , 
\ee
so $\xi_0$ does vanish as $T$. 

At high temperatures, 
both $A=g\lambda^3$ and $\lambda$ vanish. By a simple inspection of equation~(\ref{CrossOverEM}), we see that both 
$\gamma_0^*(T)$ and $\xi_0$ diverge. Using the large-$\xi$ behavior of 
$L(\xi)$, we first find $\xi_0 \sim (3 \zeta(3/2)/2)^{2/5} A^{2/5} (\gamma_0^*(T))^{6/5}$, which inserted into 
equation~(\ref{CrossOverEM}) gives
\be
\la{COHT}
\gamma_0^*(T) \sim  (5/3)^{5/4} (3 \zeta(3/2)/2)^{1/2} \frac{g^{1/2}}{\lambda} \sim {\rm cst} 
\; T^{3/4} \;\; \text{when} \;\; T \to \infty \; . 
\ee
Note that the corresponding $\xi_0$ behaves as $T^{1/2}$ and indeed diverges.

\end{document}